\documentclass[aps,prl,longbibliography,reprint,showpacs,floatfix,superscriptaddress]{revtex4-1}
\usepackage{graphicx}
\usepackage{amsmath,amssymb,amsfonts}

\begin{document}

\title{Near room temperature antiferromagnetic ordering with a potential low dimensional magnetism in AlMn$_2$B$_2$}
\author{Tej N. Lamichhane} 
\affiliation{Ames Laboratory, U.S. DOE, and Department of Physics and Astronomy, Iowa State University, Ames, Iowa 50011, USA}
\author{Khusboo Rana} 
\affiliation{Ames Laboratory, U.S. DOE, and Department of Physics and Astronomy, Iowa State University, Ames, Iowa 50011, USA}
\author{Qisheng Lin}
\affiliation{Ames Laboratory, U.S. DOE and Department of Chemistry, Iowa State University, Ames, Iowa 50011, USA}
\author{Sergey L. Bud'ko}
\affiliation{Ames Laboratory, U.S. DOE, and Department of Physics and Astronomy, Iowa State University, Ames, Iowa 50011, USA}
\author{Yuji Furukawa}
\affiliation{Ames Laboratory, U.S. DOE, and Department of Physics and Astronomy, Iowa State University, Ames, Iowa 50011, USA}  
\author{Paul C. Canfield}
\affiliation{Ames Laboratory, U.S. DOE, and Department of Physics and Astronomy, Iowa State University, Ames, Iowa 50011, USA}
%\altaffiliation{Materials Science and Technology Division, Oak Ridge National Laboratory, Oak Ridge, Tennessee 37831}
%\address[label2]{Department of Physics,University of California,Davis,CA 95616, U.S.A.}

\begin{abstract}
We present self flux growth and characterization of single crystalline  AlMn$_2$B$_2$. It is an orthorhombic (space group Cmmm), layered material with a plate like morphology. The anisotropic bulk magnetization data, electrical transport and $^{11}$B nuclear magnetic resonance(NMR) data  revealed an  antiferromagnetic (AFM) transition at 313 $\pm$ 2 K. In the magnetization data, there is also a  broad local maximum significantly above the AFM transition that could be a signature of low dimensional magnetic interactions in AlMn$_2$B$_2$. 
\end{abstract}

\maketitle
\section{Introduction}
\begin{figure}[h]
\begin{center}
\includegraphics[width=8cm]{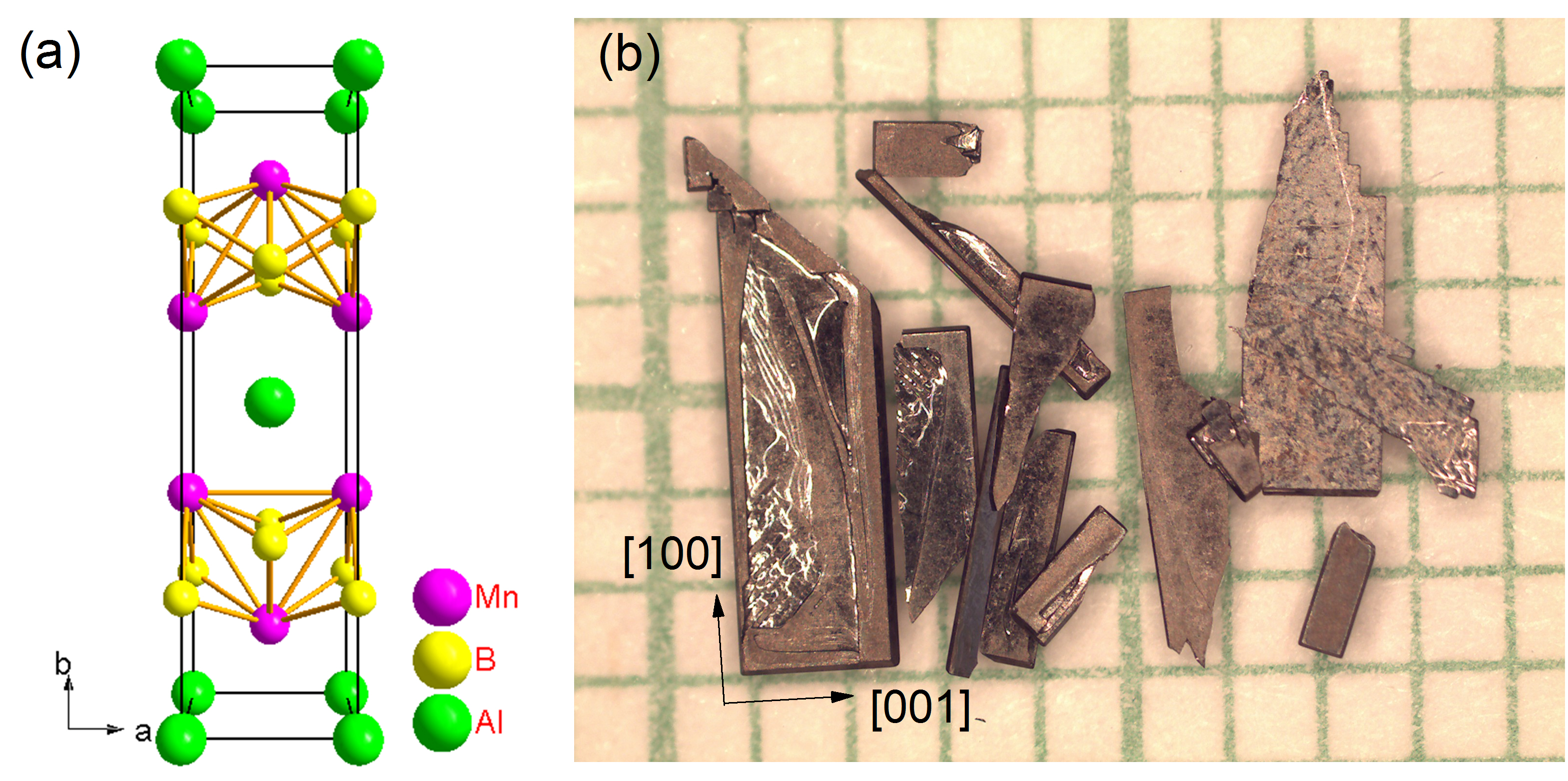}
\caption{(a)AlMn$_2$B$_2$ unit cell showing Mn$_2$B$_2$ slabs stacked with Al layer (b) Concentrated NaOH etched AlMn$_2$B$_2$ single crystals}
\label{Cryspic1}
\end{center}
\end{figure}

The AlT$_2$B$_2$ (T = Fe, Cr, Mn) system crystallizes in the orthorhombic, Cmmm structure and adopts a layer morphology with an internal structure of alternate stacking of Al atom planes and T$_2$B$_2$ slabs along the \textit{b}-axis~\cite{Du2015}. A representative unit cell of  AlMn$_2$B$_2$ is shown in Fig.~\ref{Cryspic1}(a) to demonstrate this atomic structure.  AlT$_2$B$_2$ compounds are interesting, specially for potential rare earth free magnetocaloric materials and soft magnetic materials.  AlFe$_2$B$_2$ is ferromagnetic and studied for its magneto-caloric and anisotropic magnetic properties~\cite{Tan2013, LamichhaneAlFeB2018,BARUA2018505}. Understanding the magnetic properties of the neighbouring, isostructural compounds can provide further insight in to the series as well as how to tune the magnetocaloric property of the AlFe$_2$B$_2$ via substitution. We started this work to clarify the magnetic properties of AlMn$_2$B$_2$ since it was identified as a nonmagnetic material~\cite{CHAI201552}. In addition, some inconsistencies between bulk and local probe magnetic measurements in the Al(Fe$_{1-x}$Mn$_x$)$_2$B$_2$ were observed. A later  first principle calculation suggested that  AlMn$_2$B$_2$ should be an anti-ferromagnetic compound~\cite{KeLiqin2017}.  In a recent powder neutron study, AlMn$_2$B$_2$ is identified as a ceramic AFM compound~\cite{POTASHNIKOV2019468} with Neel temperature around 390 K. A study of lattice parameters variation from room temperature to 1200 K revealed that there is a  change in anisotropy nature in \textit{a} and \textit{c} lattice parameters around 450 K and a local minimum in \textit{b} lattice parameters around 400 K~\cite{Verger2018}. The lack of a clear description of the nature or number of magnetic phase transitions in AlMn$_2$B$_2$ led us to grow and systematically study single crystalline samples.

%The lack of coherent information in all these available data provided us enough enthusiasm for the systematic study of magnetic properties of AlFe$_2$B$_2$.   

This paper reports the synthesis of bulk single crystals via high-temperature solution growth and their characterization via high and low temperature magnetization, NMR, and electrical resistance measurements. We find that AlMn$_2$B$_2$ is a metallic antiferromagnet with a transition temperature of \textit{T}$_N$ = $313\pm2$ K. In addition we find that    
AlMn$_2$B$_2$ has features associated with pseudo-two-dimensional magnets.

\begin{table}
\begin{center}
\caption{\label{tbl:crystaldata}Crystal data and structure refinement for AlMn$_{2}$B$_{2}$.}
\begin{tabular}{|l|l|}
\hline
Empirical formula & AlMn$_{2}$B$_{2}$ \\
Formula weight & 158.48 \\
Temperature & $296(2)$~K \\
Wavelength & $0.71073$~\AA \\
Crystal system, space group & Orthorhombic,  \textit{Cmmm} \\
Unit cell dimensions & a=2.9215(1)~\AA \\
 & b = 11.0709(6)~\AA \\
 & c = 2.8972(2)~\AA \\
Volume &  93.706(9) $10^3$ \AA$^3$ \\
Z, Calculated density & 2,  5.63 g/$cm^3$ \\
Absorption coefficient & 6.704  mm$^{-1}$ \\
F(000) & 73 \\
$\theta$ range ($^\circ$) & 3.693 to  29.003 \\
Limiting indices & $-5\leq h \leq5$\\
& $-22\leq k\leq 22$\\
& $-5\leq l \leq 5$ \\
Reflections collected & 1467\\
Independent reflections & 270 [R(int) = 0.0401] \\
Completeness to theta = 25.242$^{\circ}$ & 98.5\% \\
Absorption correction & multi-scan, empirical \\
Refinement method & Full-matrix least-squares \\
Data / restraints / parameters & 270 / 0 / 12 \\
Goodness-of-fit on $F^2$ & 1.101 \\
Final R indices [I$>2\sigma$(I)] & $R1 = 0.0362$, $wR2 = 0.0817$ \\
R indices (all data) & $R1 = 0.0387$, $wR2 = 0.0824$\\
Largest diff. peak and hole & 2.341  and -1.249 e.\AA$^{-3}$\\ \hline
\end{tabular}
\end{center}
\end{table}

\begin{table}
\begin{center}
\caption{\label{tbl:atomcoord}Atomic coordinates and equivalent isotropic displacement parameters (A$^2$) for AlMn$_{2}$B$_{2}$. U(eq) is defined as one third of the trace of the orthogonalized U$_{ij}$ tensor.}
\begin{tabular}{| p{0.6cm} | p{1.2cm} | p{1.2cm} | p{1.8cm} | p{1.2cm} |p{1.8cm}|}
\hline
atom & Wyckoff site & x & y & z & U$_{eq}$ \\ \hline
Mn & 4\textit{j} & 0 & 0.3552(1) & 1/2 & 0.0070(1) \\ \hline
Al & 2\textit{a} & 0 &  0 & 0 & 0.0067(5) \\ \hline
B & 4\textit{i} &  0 & 0.2065(5) & 0 & 0.0070(1) \\ \hline
\end{tabular}
\end{center}
\end{table}

\begin{figure}
\begin{center}
\includegraphics[width=8cm]{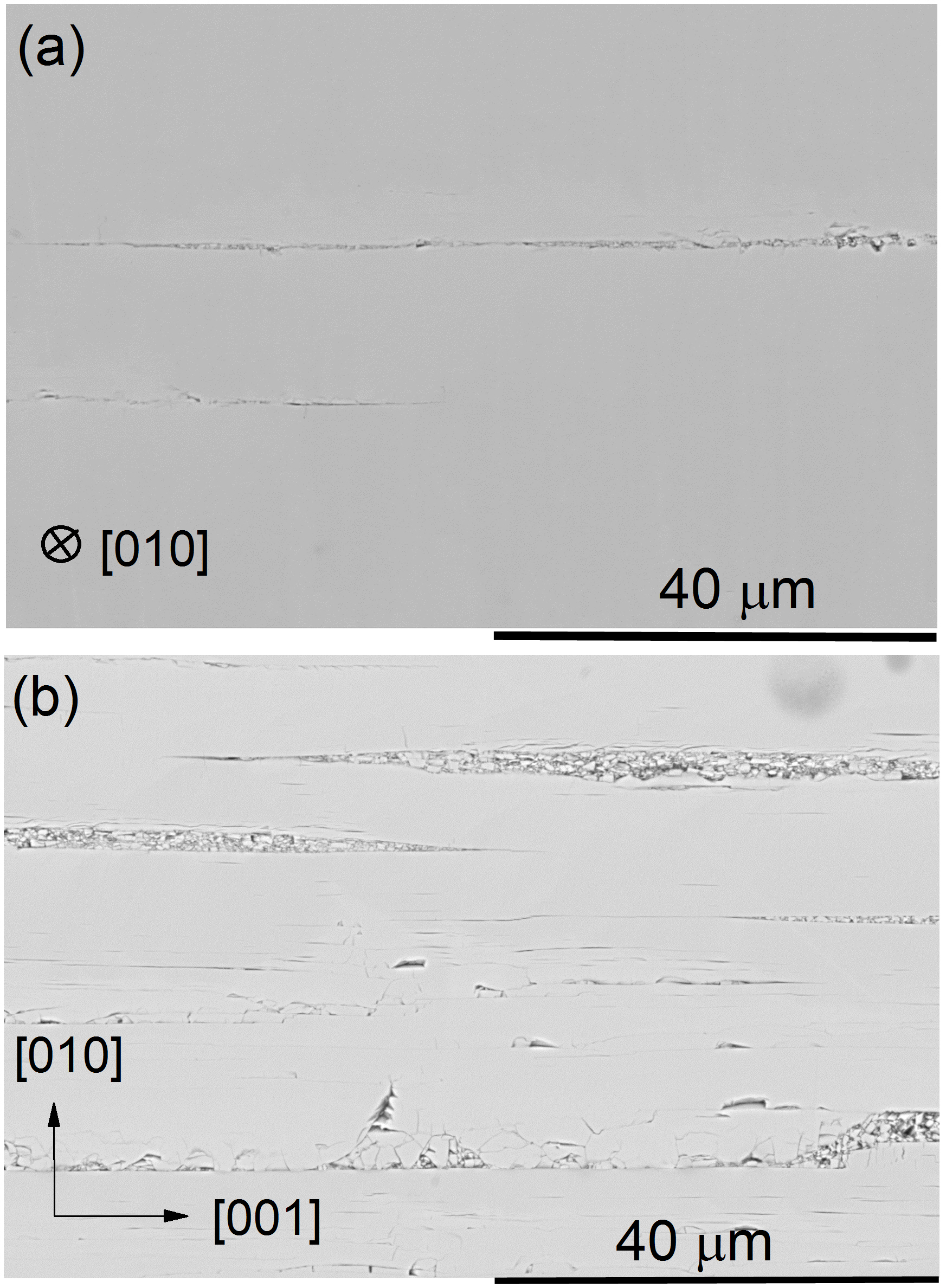}
\caption{(a) SEM image of AlMn$_2$B$_2$ single crystalline sample along the planar view  (with electron beam parallel to [010]) (b) SEM image of AlMn$_2$B$_2$ in a cross sectional view with electron beam parallel to [100].}
\label{Cryspic2}
\end{center}
\end{figure}

\begin{figure}[!ht]
\begin{center}
\includegraphics[width=9.5cm]{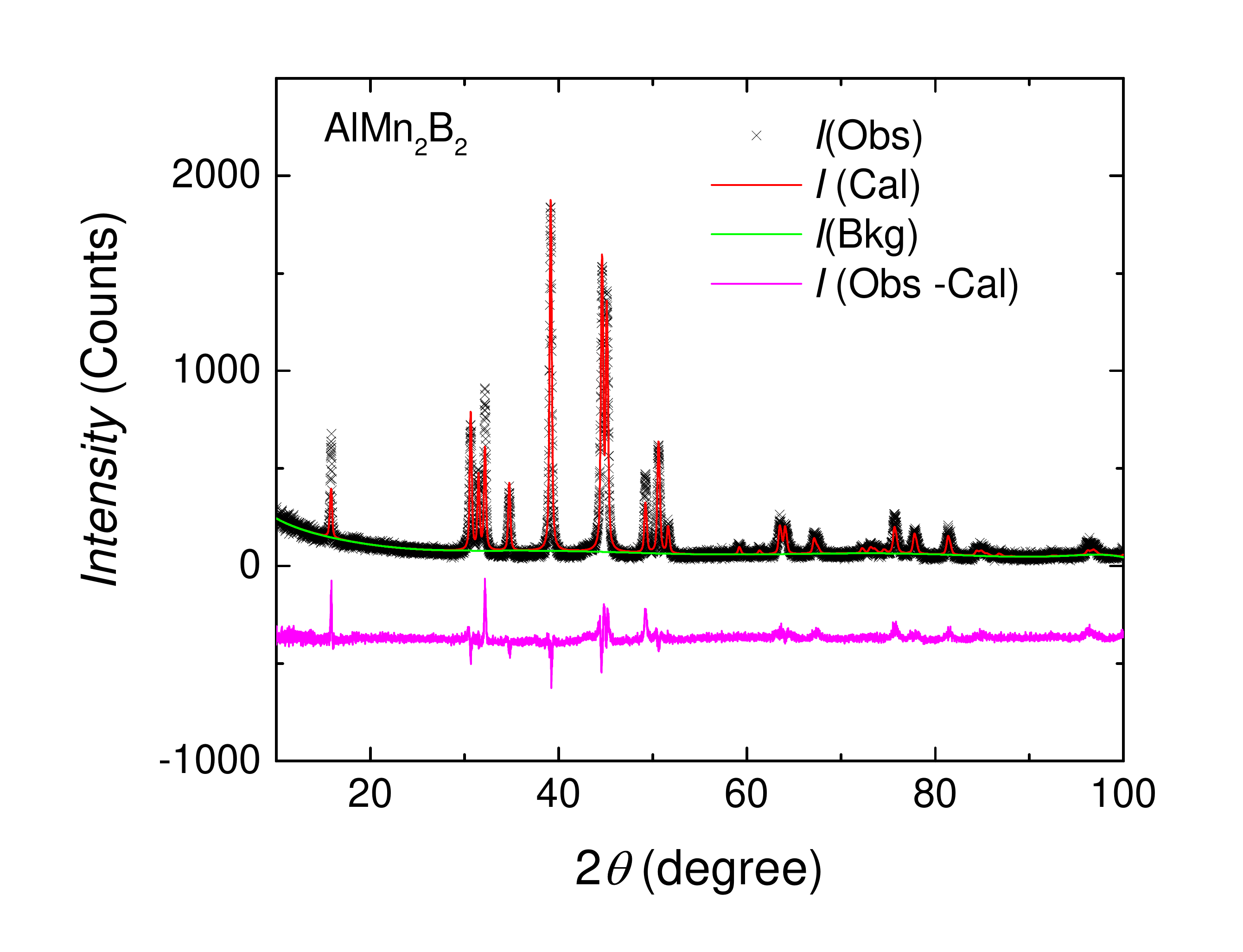}
\caption{Single crystal crushed powder XRD pattern where \textit{I }(Obs), \textit{I}(Cal), \textit{I}(Bkg) and \textit{I}(Obs-Cal) are observed, calculated, background and differential diffractrograms respectively.}
\label{TexturedXRD}
\end{center}
\end{figure}
   
\begin{figure}[!ht]
\begin{center}
\includegraphics[width=8cm]{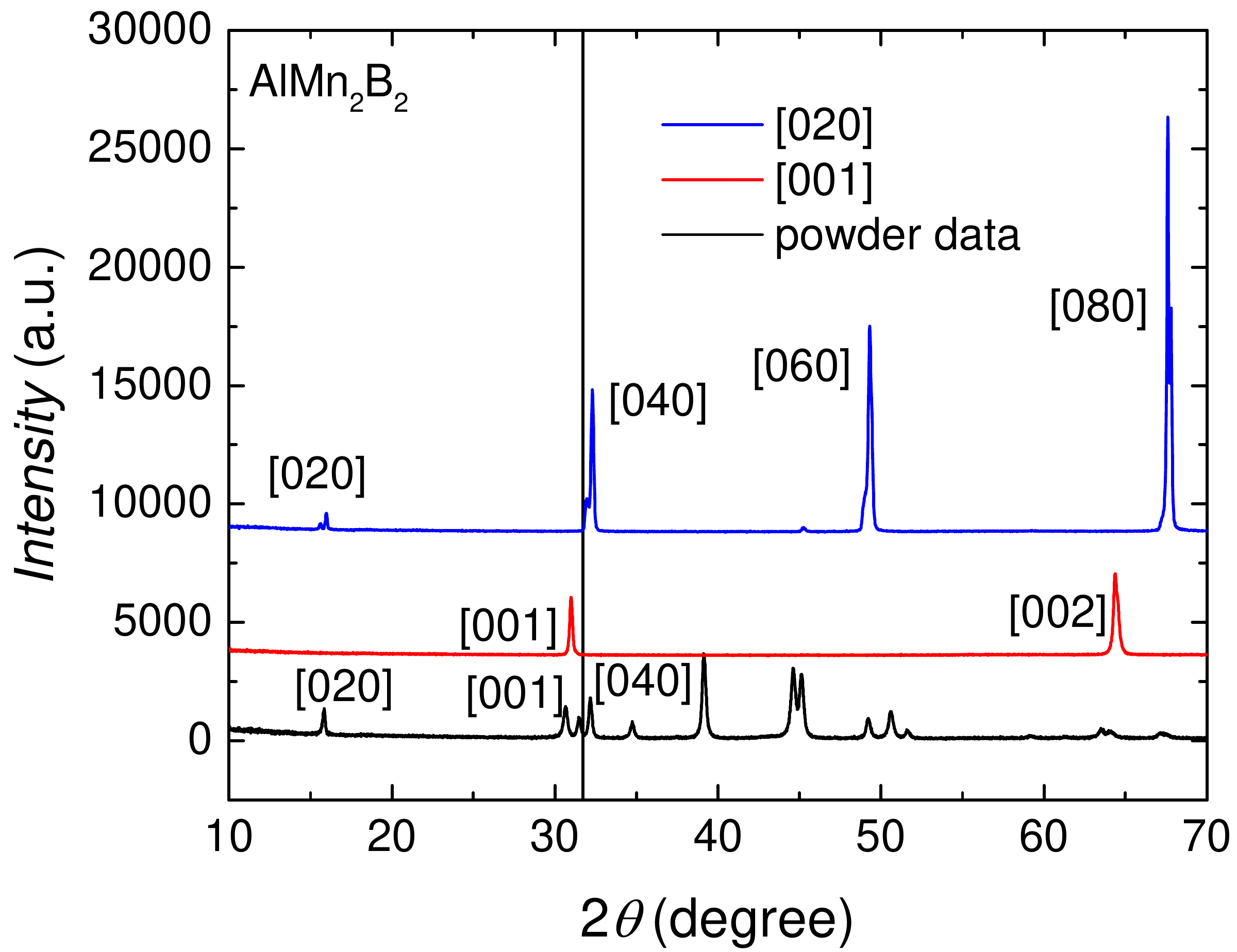}
\caption{Crystallographic orientation characterization of AlMn$_2$B$_2$ surfaces using monochromatic Cu \textit{K}$_{\alpha}$ radiation in Bragg Brentano diffraction geometry. The top curve shows the family of [020] peaks identifying direction perpendicular to plate as [010]. The direction along the thickness of the plates is found to be [001] leaving the direction along the length as [100]. The vertical grid line through the [110] powder diffraction peak (not labeled in diagram) is a reference to identify [001] and [040]  peaks observed for different facets.}
\label{SCXRD}
\end{center}
\end{figure}

\section{Experimental Details}
\subsection{Crystal growth}

Solution growth is a powerful tool even for compounds with high melting elements like B~\cite{Belashchenko2015,LamichhaneAlFeB2018,Canfield2001}. The major difficulty associated with solution growth is finding an initial composition that allows for growth of the single phase, desired compound. For example, CaKFe$_4$As$_4$ growth in single phase form presents an illustrative example~\cite{Meier2017}. Fortunately, with the innovation of fritted alumina crucibles sets~\cite{Canfieldfrittedcrucible2016} we can now reuse decanted melt and essentially fractionate the melt, as described below. 

Al shot (Alfa Aesar 99.999\%), B pieces (Alfa Aesar 99.5\% metal basis) and Mn pieces (Alfa Aesar 99.9\% metal basis) after surface oxidation cleaning as described elsewhere~\cite{LamichhaneZrMnP} were used for the crystal growth process. We started with an Al rich composition, Al$_{68}$Mn$_{22}$B$_{10}$, and arc-melted it at least 4 times under an Ar atmosphere. The button was then cut with a metal cutter and re-arcmelted if some not-reacted B pieces were found. After the button appeared to be homogeneous, it was packed in a fritted alumina crucible set~\cite{Canfieldfrittedcrucible2016} and sealed under partial pressure of argon inside amorphous silica jacket to form a growth ampoule. The growth ampoule was then heated to 1200~$^\circ$C  over 2 h and soaked there for 10 h before spinning using a centrifuge. Due to high melting point of B containing compounds, homogeneous liquid was not formed at 1200~$^\circ$C.  Undissolved polycrystalline  MnB and  Al-Mn binary compounds were separated at 1200~$^\circ$C via centrifuging. The catch crucible collected the homogeneous melt at 1200~$^\circ$C was again sealed in a fritted alumina crucible sets under Ar atmosphere to form second growth ampoule. This second ampoule was heated to 1200~$^\circ$C over 2 h, held there for another 10 h and cooled down to 1100~$^\circ$C over 50 h and spun using centrifuge to separate the crystals. The second growth attempt produced a mixture of the targeted AlMn$_2$B$_2$ phase along with MnB crystals. So as to avoid this MnB contamination, the catch crucible of the second growth was used for a third growth and sealed again under a partial pressure of Ar. For this stage, to make sure there are no other nucleated crystals, the third growth growth was heated to 1200~$^\circ$C over 2 h and soaked there for 2 h. It was then cooled down to 1100~$^\circ$C over 1 h and stayed there for 1 h followed by slow cooling to 990~$^\circ$C over 120 h and centrifuged to separate large, single phased AlMn$_2$B$_2$ crystals as shown in Fig.~\ref{Cryspic1}(b). The flux on the surface was removed via concentrated NaOH etching.

It should be noted that predominantly single phase AlMn$_2$B$_2$ crystals were grown in single growth attempt using initial Al$_{84}$Mn$_{8}$B$_{8}$ composition however the crystals were small, due to multiple nucleation sites.

%To get the millimeters sized crystals in a single step, we need to go higher than 1200~$^{\circ}$C in a vertical tube furnace which is much greater work than recycling the fractionanted melt.

\section{Crystal structure and stoichiometry}
As grown single crystals were characterized using a scanning electron microscope (SEM), as well as both powder and single crystal X-ray diffraction (XRD). Figures~\ref{Cryspic2}(a) and (b) show the planar and cross sectional backscattered SEM images of AlMn$_2$B$_2$ single crystals which show predominantly homogeneous compositions. The small linear grooves are the cracked layers associated with the SEM sample polishing. Being a layered material, it can be easily cleaved and deformed. Boron is difficult to account for correctly in electron dispersive spectroscopy(EDS), as a consequence of this we determined only the Mn:Al ratio for two different batches of single crystalline samples. In first batch, 13 spots were analyzed in EDS with Mn:Al ratio of 2.07 for all characteristics X-ray emissions. Similarly, an 8 spot analysis in the second batch provided the Mn:Al ratio to be 2.12 for characteristics \textit{K}-lines for all elements. With the \textit{L}-characteristics-lines analysis, a ratio of 2.51 was obtained for the second batch. Without the creation and use of Mn-Al-B based standards, further characterization by EDS is difficult.

%These results suggest that, the accuracy of the EDS characterization of B containing sample depends on sample sample conditions, analyzing techniques and demands some kind of complementary characterization.

Although the EDS results are qualitatively in agreement with the AlMn$_2$B$_2$ structure, to more precisely determine the composition and structure, multiple batches of AlMn$_2$B$_2$ were investigated using single crystal XRD technique. Single crystalline XRD data were collected with the use of graphite monochromatized Mo \textit{K}$_{\alpha}$ radiation ($\lambda$=0.71073 \AA) at room temperature on a Bruker APEX2 diffractometer. Reflections were gathered by taking five sets of 440 frames with 0.5$^\circ$ scans in $\omega/\theta$, with an exposure time of 10 s per frame and the crystal-to-detector distance of 6 cm. The structure solution and refinement for single crystal data was carried out using SHELXTL program package. Attempts to refine occupancies of each site indicated full occupancy($<3\sigma$). The final stage of refinement was performed using anisotropic displacement parameters for all the atoms. The refinement metrics and atomic coordinates are presented in TABLE~\ref{tbl:crystaldata} and~\ref{tbl:atomcoord} respectively. The single crystalline refinement showed AlMn$_2$B$_2$ as a stoichiometric material.

Etched single crystals were finely ground and spread over a zero background silicon wafer sample holder with help of a thin film of Dow Corning high vacuum grease. Powder diffraction data were obtained using a Rigaku Miniflex II diffractometer within a 2$\theta$ range of 10 - 100$^\circ$ with a step of 0.02$^\circ$ and dwelling time of 3 seconds for data acquisition. The crystallographic information file from the single crystal XRD solution was used to fit the powder XRD data using GSAS~\cite{Larson2004} and EXPGUI~\cite{Toby2001JAC} software packages. Figure~\ref{TexturedXRD} shows the Rietveld refined powder XRD pattern with R factor of 0.08. Being a relatively hard, layered  material, texture is visible along [020] direction although March Dollase texture correction was employed to account for this intensity mismatch. 

To identify the crystallographic orientation of the AlMn$_2$B$_2$ single crystals, we employed the monochromatic X-ray diffraction from the crystallographic surfaces in the Bragg-Brentano geometry ~\cite{AJesche2016,LamichhaneAlFeB2018}. The direction perpendicular to the plate was identified to be [010] since a family of \{020\} lines were observed in the diffraction pattern as shown in blue curve in Fig.~\ref{SCXRD}. The plate was held vertical and the family of \{001\} peaks were obtained as shown in red curve of Fig.~\ref{SCXRD}. The monochromatic x-ray surface diffraction peaks were compared with powder diffraction data to correctly identify their directions. A vertical line through the powder [110] peak was used as a reference point of comparison as shown in Fig.~\ref{SCXRD}. Then the last remaining direction was identified to be [100] along the length of the crystals. A reference coordinate system is shown in Fig.~\ref{Cryspic1}(b) to demonstrate the crystallographic orientations of AlMn$_2$B$_2$ crystals.

\begin{figure}[!ht]
\begin{center}
\includegraphics[width=8cm]{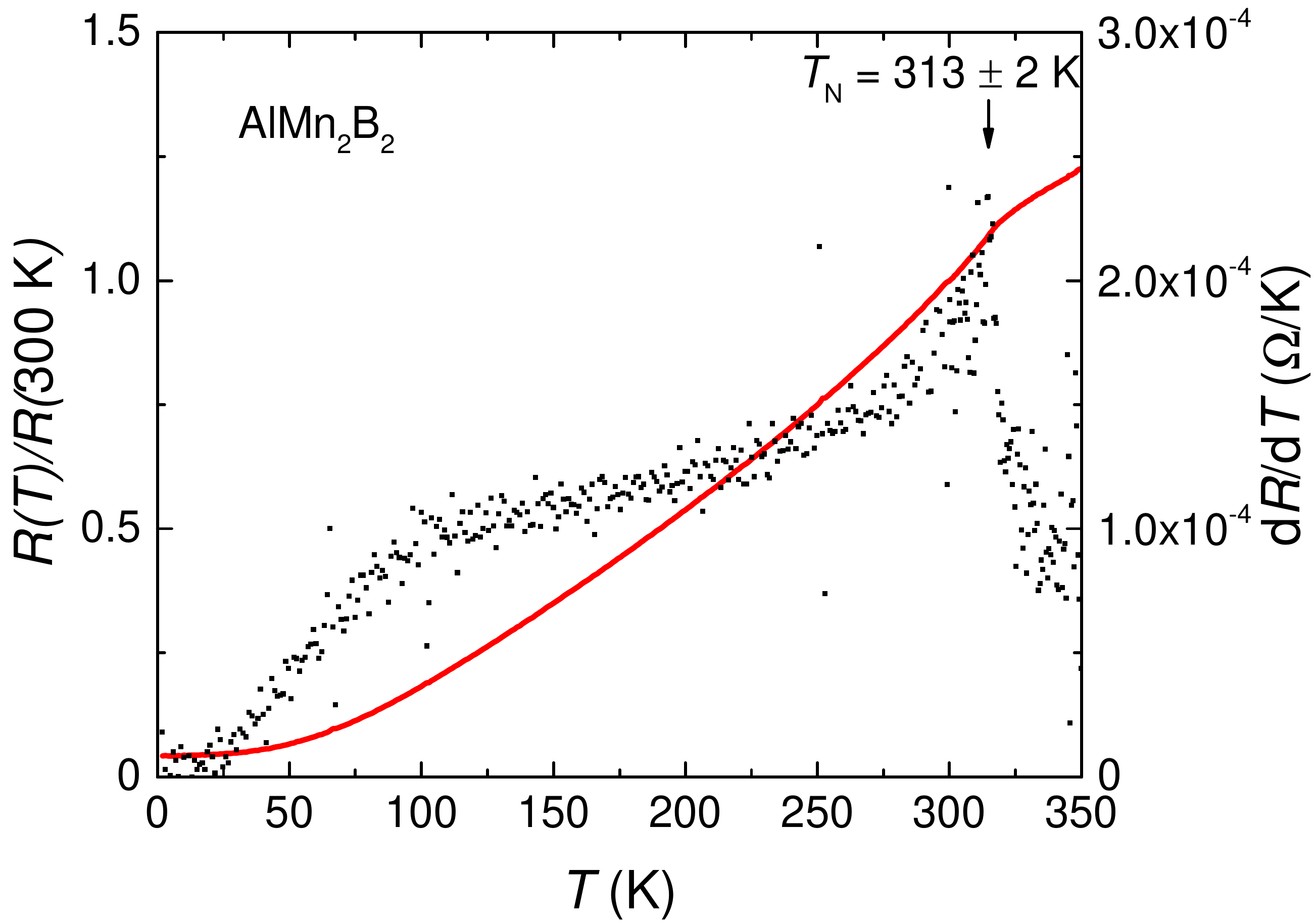}
\caption{Temperature dependent normalized resistance (left axis) and temperature derivative (right axis) of AlMn$_2$B$_2$. The resistance is metallic in nature. The temperature derivative shows an anomaly at 313 $\pm$ 2 K consistent with an AFM phase transition.}
\label{Resistivityandderivative}
\end{center}
\end{figure}

\begin{figure}[!ht]
\begin{center}
\includegraphics[width=8cm]{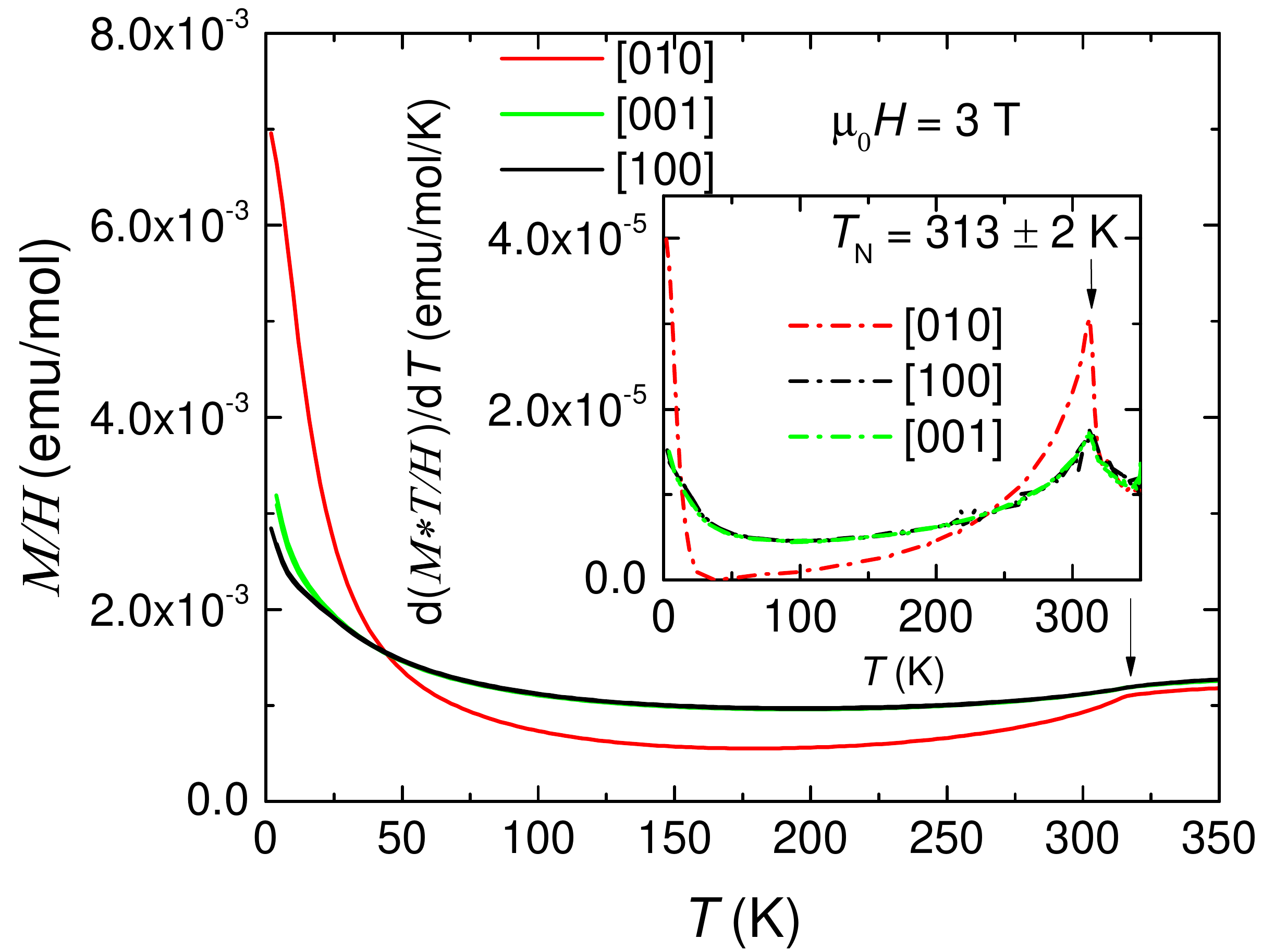}
\caption{Low temperature (2 - 350 K) \textit{M/H} along various crystallographic axes of AlMn$_2$B$_2$ sample as outlined in the graph. The inset shows $\frac{d(M*T/H)}{dT}$ as a function of temperature.}
\label{MPMSMTanddkidtinset}
\end{center}
\end{figure}

\begin{figure}[!b]
\begin{center}
\includegraphics[width=8cm]{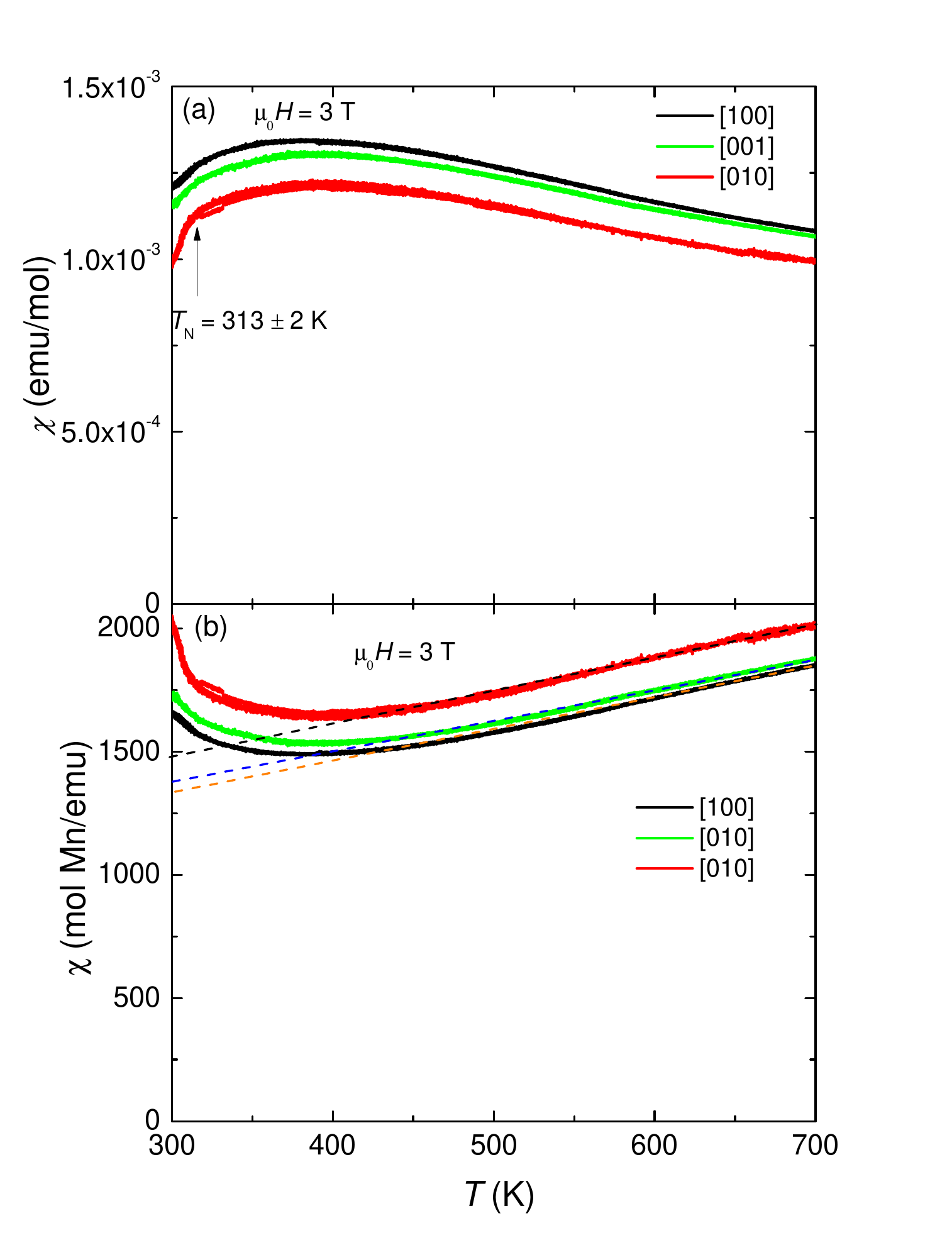}
\caption{(a) High temperature susceptibility data along various axes measured using VSM. There are shallow anomalies present around 313 $\pm$ 2 K for each directions. (b) Corresponding Curie Weiss plots identifying AlMn$_2$B$_2$ as an AFM material with $\theta_{010}$ = - 815 K, $\theta_{100}$ = - 750 K, and $\theta_{001}$ = - 835 K respectively.}
\label{CurieWeissOnly}
\end{center}
\end{figure}

\begin{figure}[!b]
\begin{center}
\includegraphics[width=8cm]{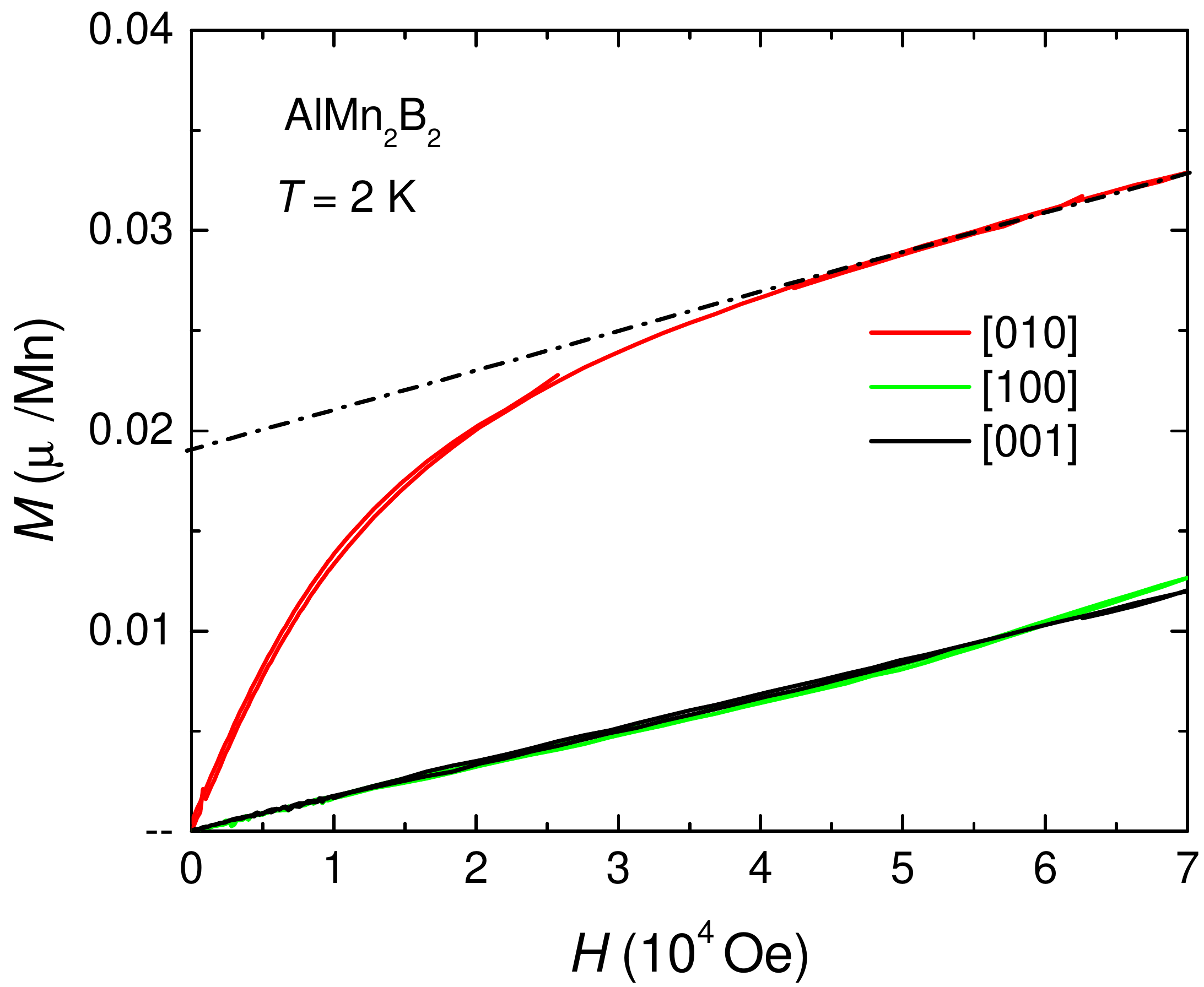}
\caption{Field dependent magnetization \textit{M(H)} of AlMn$_2$B$_2$ at 2 K. The magnetization along [010] direction shows a saturation magnetization of ~ 0.02 $\mu_B$/Mn with respect to other two principle directions outlined with a linear fit of the high field region data. The \textit{M(H)} data along [010] direction shows no magnetic hysteresis i.e. the almost overlapping 2 red curves for increasing and decreasing field. The At higher field region, all 3 \textit{M(H)} data have the same slopes.}
\label{MH}
\end{center}
\end{figure}
 
\section{Electric and Magnetic properties}

The temperature dependent electrical resistance of AlMn$_{2}$B$_{2}$ was measured in a traditional 4 probe measurement on a NaOH etched, rod like sample using an external device control option to interface with a Linear Research, Inc. ac (1 mA, 17 Hz) resistance bridge (LR 700). Thin platinum wires were attached to the sample using DuPont 4929N silver paint to make electrical contact. Quantum Design Magnetic Property Measurement System (MPMS) was used as a temperature controller. The measured temperature dependent electrical resistance of AlMn$_{2}$B$_{2}$ is shown in Fig.~\ref{Resistivityandderivative}. These data further confirm that our single crystals are essentially stoichiometric AlMn$_{2}$B$_{2}$; given that the residual resistivity ratio ($\frac{R(350.0 K)}{R(2.0 K)}$) is 28.5, there is relatively low disorder scattering.  In addition, a very clear feature is seen in both R(T) and $\frac{d(R(T))}{dT}$ at T = 313 $\pm$ 2 K. Such features are often related to a loss of spin disorder scattering at a magnetic transition~\cite{MFisher1968}.  As such, these data are our first suggestion that AlMn$_{2}$B$_{2}$ may indeed have some form of magnetic order below 315 K.

The magnetic properties of AlMn$_{2}$B$_{2}$ were studied from a base temperature of 2 K to 700 K. Low temperature anisotropic magnetization data of single crystalline AlMn$_{2}$B$_{2}$ samples were measured within the temperature range 2 - 350 K using a MPMS. High temperature, anisotropic temperature dependent magnetization data were obtained using a Quantum Design VersaLab Vibrating Sample Magnetometer (VSM) over the  temperature range 300 - 700 K in an oven option mode. 

The low temperature anisotropic susceptibility data, with \textit{H} = 3 T applied field, are presented in Fig.~\ref{MPMSMTanddkidtinset}. Below 50 K, the magnetization data show a low temperature  upturn as reported in a previous literature~\cite{CHAI201552}. In all three directions, there is a clear anomaly in susceptibility data around 312 K. The inset shows $\frac{d(M*T/H)}{dT}$ as a function of temperature~\cite{MFishercriterion} showing a clear anomaly around 312 K identifying AlMn$_{2}$B$_{2}$ as a AFM material. The observed anomaly in $\frac{d(M*T/H)}{dT}$  coincides with the kink observed in $\frac{dR}{dT}$.  

%A relevant information available to AlMn$_2$B$_2$ around this temperature is an in-situ XRD study which reports a switching in magnitude of \textit{a} and \textit{c} lattice parameters around 400 K along with a local minimum in b lattice parameter~\cite{Verger2018}. Similar variations in lattice parameters and measured magnetic moment were studied using powder neutron diffraction data but the features were little different temperature $\sim 470$ K. Potashnikov \textit{et} \textit{al.} attributed that these change in lattice parameters could be related to magnetorestriction phenomenon~\cite{POTASHNIKOV2019468}.

Recently, AlMn$_{2}$B$_{2}$ was reported to be AFM however Neel temperature was reported to be around 390 K~\cite{POTASHNIKOV2019468}. To examine higher temperatures, our high temperature susceptibility data, obtained using our VSM are presented in Fig.~\ref{CurieWeissOnly}(a) and (b). Although a broad local maximum of the susceptibility  around 350 - 390 K for different axes, consistent to reference[\onlinecite{POTASHNIKOV2019468}] was found, the  $\frac{d(M*T/H)}{dT}$ did not show any anomaly.  The only clear and conclusive feature in the high temperature data associated with a magnetic transition is the feature at 313 $\pm$ 2 K. The broad local maximum in magnetization well above the transition temperature can be associated with low dimensional, or linear chain anisotropic Heisenberg anitiferromagnetism~\cite{JMa2013,VASILEV20001619,Kim1998,Bonner1964,Dingle1969}. The fitted Curie Weiss temperatures for various axes were obtained to be $\theta_{010}$ = - 815 K, $\theta_{100}$ = - 750 K, and $\theta_{001}$ = - 835 K. From the average slope of Curie Weiss plot, the effective moment of Mn is found to be $\sim$2.5$\mu_B$/Mn.

At low temperature, T $\leq$ 50 K, in Fig.~\ref{MPMSMTanddkidtinset} there is a clear upturn in the \textit{M/H} data, particularly for \textit{H} along the [010] direction. In order to better understand this we measured the anisotropic field dependent magnetization at 2 K as shown in Fig.~\ref{MH}. For fields greater than 4 T the slopes of the \textit{M(H)} plots are comparable for all three directions. For \textit{H} $\parallel$ [010], there is a roughly 0.02 $\mu_B$/Mn offset due to a rapid increase and saturation for \textit{H} $\leq$ 3 T. The origin of this small, anisotropic contribution is currently not known.

\begin{figure}
\begin{center}
\includegraphics[width=8cm]{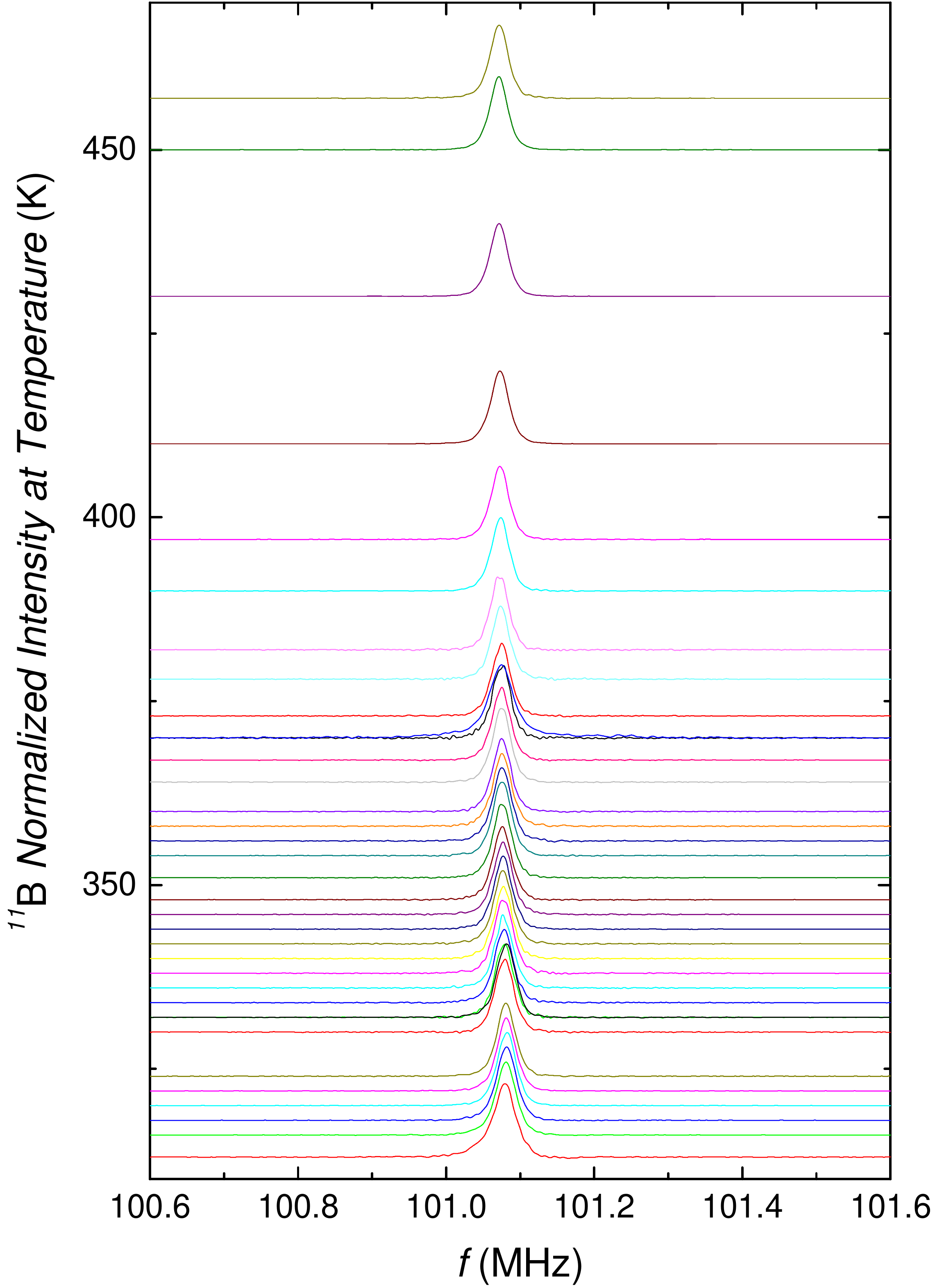}
\caption{ $^{11}$B NMR spectrum measured at different temperatures from 315 to  457 K with H = 7.4089 T.}
\label{11BNMRspectrum315-457K}
\end{center}
\end{figure}

\begin{figure}
\begin{center}
\includegraphics[width=8cm]{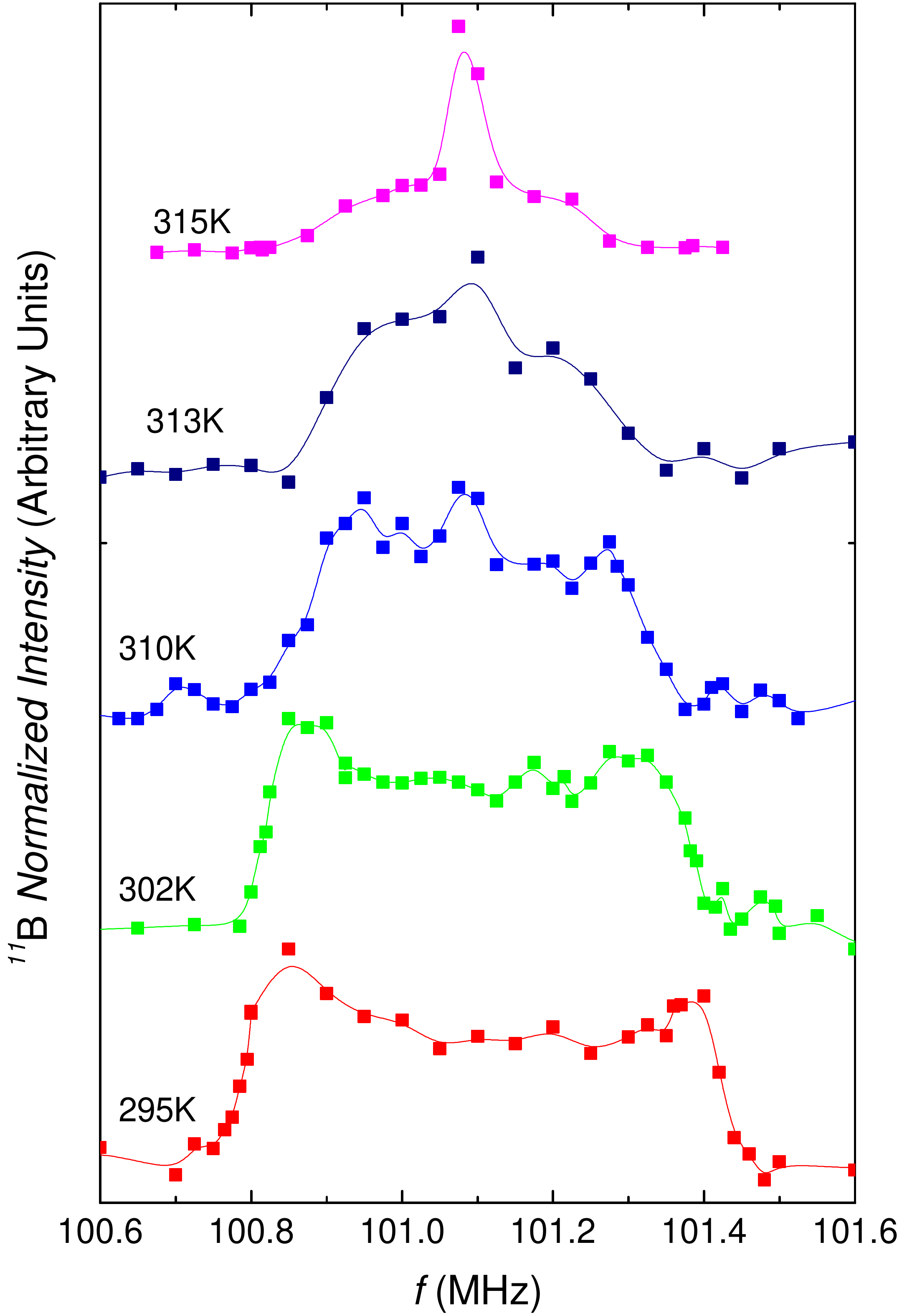}
\caption{ $^{11}$B NMR spectra measured at H = 7.4089 T by sweeping frequency.}
\label{11BNMRspectrum295-315K}
\end{center}
\end{figure}

\begin{figure}[!b]
\begin{center}
\includegraphics[width=8cm]{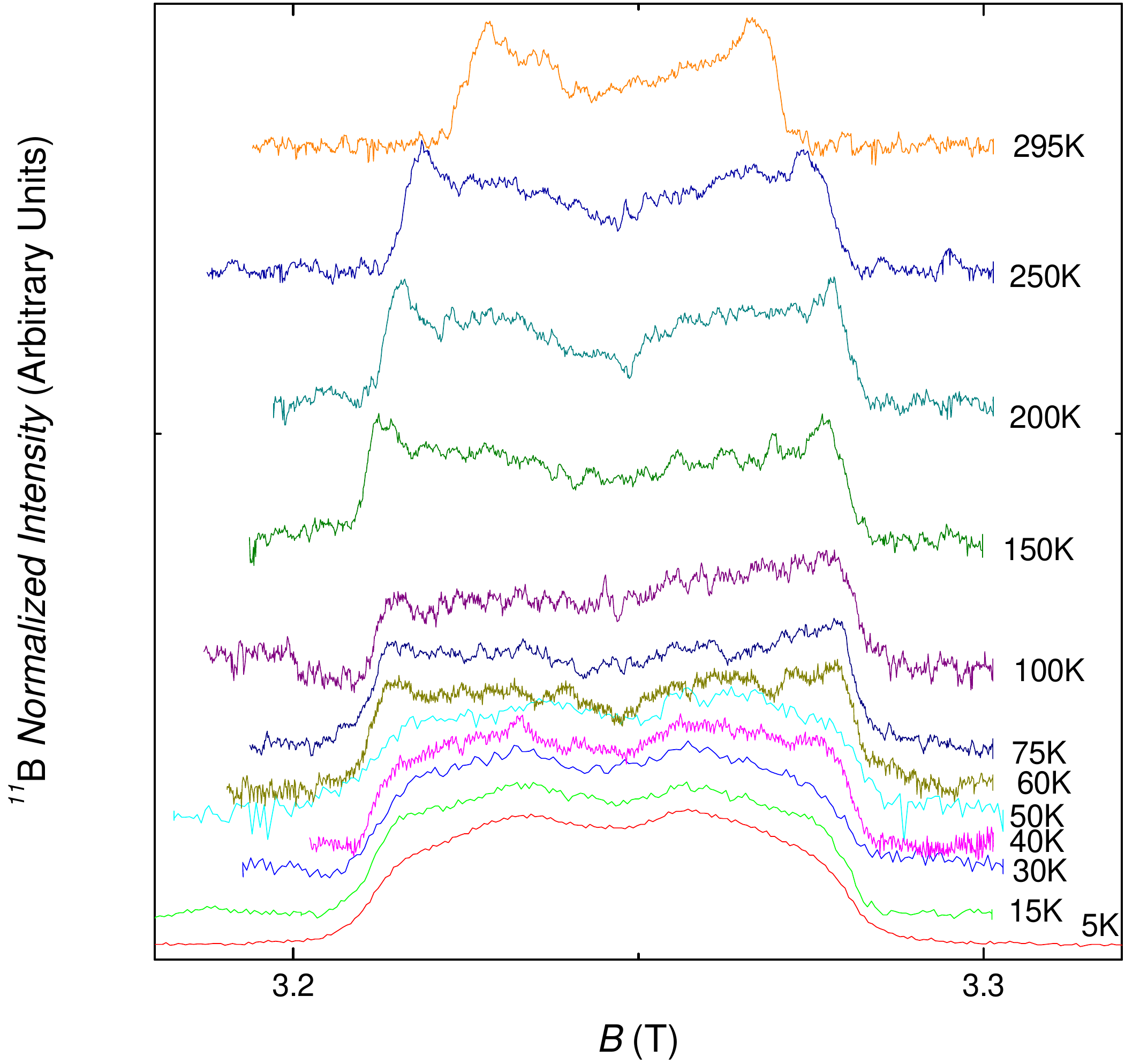}
\caption{ $^{11}$B NMR spectra measured at different temperatures between 5 K and 295 K measured using field sweeping method. A noticeable change in shape of $^{11}$B NMR peaks around 50 K coincides with changing the magnetic anisotropy between [100]/[001] and [010] directions as shown in Fig.~\ref{MPMSMTanddkidtinset}.}
\label{11BNMRspectrum5-295K}
\end{center}
\end{figure}

\begin{figure}
\begin{center}
\includegraphics[width=8cm]{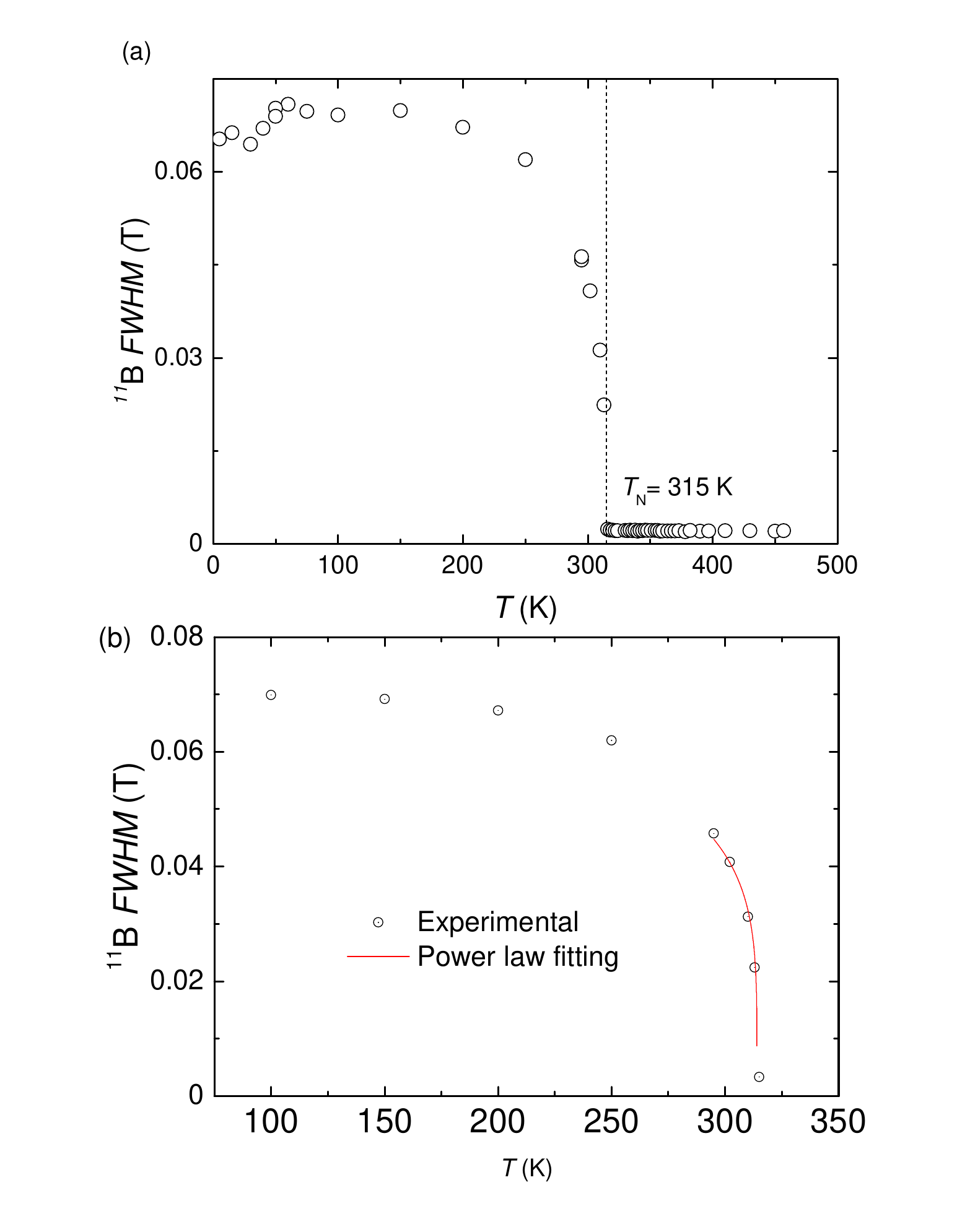}
\caption{(a) Temperature dependence of FWHM $^{11}$B NMR spectra in powdered AlMn$_{2}$B$_{2}$ sample showing AFM transition around 315 K.  (b) Power law fitting of the observed temperature variation of  $^{11}$B FWHM in the temperature range  295 - 315 K as $FWHM \propto [1-(\frac{T}{T_{\rm N} })]^{\beta} $ with $T_{\rm N}$ = 314 K and $\beta = 0.21 \pm 0.02$.}
\label{TempvariationNMRparameter}
\end{center}
\end{figure}

\begin{figure}[!ht]
\begin{center}
\includegraphics[width=8cm]{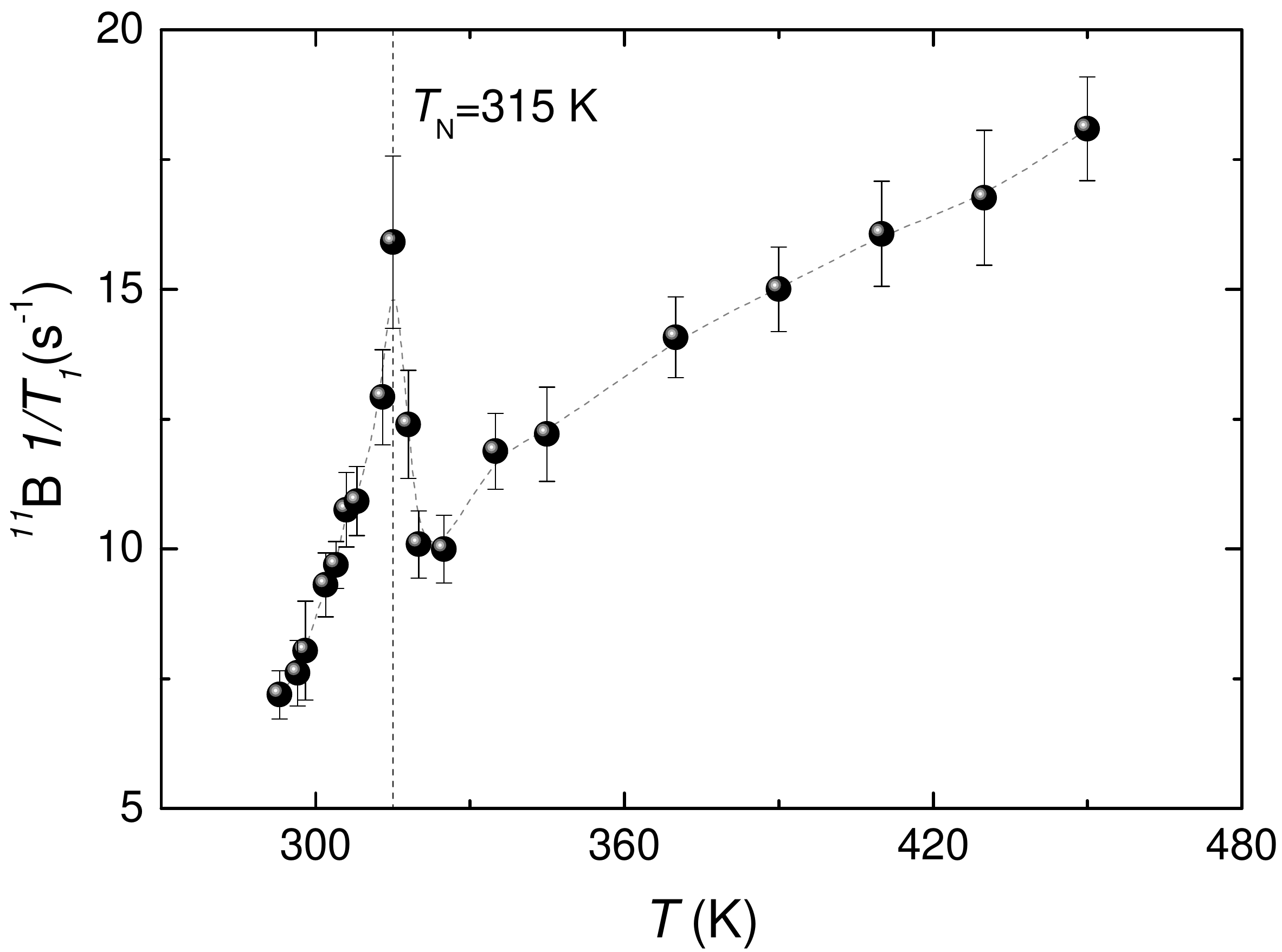}
\caption{The inverse of relaxation rate (\textit{T}$_1$) is plotted as a function of \textit{T} from 293 K to 450 K. The transition temperature at 315 K is evidenced by a sharp peak of $\frac{1}{T_1}$.}
\label{Temperaturevariationof1overT}
\end{center}
\end{figure}

\section{Nuclear Magnetic resonance Study}
%Though the Fisher's criterion on magnetization data and resistivity revealed a clear AFM transition, the low temperature anomaly below 50 K and high temperature broad peak around 390 K were not clearly understood. To investigate all these anomalies  and the transition around 312 K, 

To further investigate the magnetism of AlMn$_2$B$_2$, we carried out  $^{11}$B NMR measurements at various temperatures between 5 K and 457 K as presented in Figs. ~\ref{11BNMRspectrum315-457K} - ~\ref{11BNMRspectrum5-295K}. To perform the NMR measurements for the temperature region of \textit{T} = 5 - 295 K, crushed single crystalline powder was enclosed in a weighing paper folded closed cylindrical tube and inserted inside the NMR coil. For the higher temperature NMR measurements up to 457 K from room temperature, the crushed powder was sealed under $\frac{1}{3}$ atmospheric pressure of Ar inside a $\sim$1 mm internal diameter amorphous silica tube. The NMR measurements were carried out using a lab-built phase coherent spin-echo pulsed NMR spectrometer  on $^{11}$B  (nuclear spin \textit{I} = $\frac{3}{2}$ and gyromagnetic ratio $\frac{\gamma_N}{2 \pi}$ = 13.6552 MHz/T) nuclei in the temperature range  5 $<$ T $<$ 457 K. NMR spectra were obtained either by Fourier transform of the NMR echo signals, by sweeping frequency  or by sweeping magnetic field. Magnetic phase transition was studied analyzing the full width at half maximum (FWHM) of $^{11}$B NMR spectra and spin-lattice relaxation rate $\frac{1}{T_1}$. The $^{11}$B $\frac{1}{T_1}$ was measured by the conventional single saturation pulse method.

Figure~\ref{11BNMRspectrum315-457K} shows the $^{11}$B NMR spectra obtained by Fourier transform of the NMR spin echo for temperatures in the range 315 - 457 K at H = 7.4089 T. Throughout the range of study, the FWHM $\sim$29 kHz is nearly independent of temperature. On the other hand, below 315 K, as shown in Fig.~\ref{11BNMRspectrum295-315K}, the $^{11}$B NMR line broadens abruptly and has an almost rectangular shape at low temperatures. Since the rectangular shape is characteristic of NMR spectrum in AFM ordered state for powder sample, the results clearly indicate that the magnetic phase transition around 315 K is AFM. Similar rectangular NMR spectra in AFM state have been observed in BiMn$_2$PO$_6$~\cite{Nath2014}, NaVGe$_2$O$_6$~\cite{Pedrini2004}, CuV$_2$O$_6$~\cite{Kikuchi2000}, and BaCo$_2$V$_2$O$_8$~\cite{Ideta2012} . 

%Interestingly this anomaly in $^{11}$B FWHM coincides with anomaly in resistance as shown in Fig.~\ref{Resistivityandderivative} and  magnetic susceptibility as shown Fig.~\ref{MPMSMTanddkidtinset}. All of these results clearly suggest an antiferromagnetic transition near 315 K. \\*

%although there is a broad local maximum in the magnetic susceptibility data as shown in Fig.~\ref{CurieWeissOnly}(a). For comparison, similar frequency sweep data are presented in Fig.~\ref{11BNMRspectrum295-315K} in some selective temperatures in the temperature range 295 K to 315 K. Below 315K,

In the low temperature range between 5 - 295 K, several $^{11}$B NMR spectra were measured at a frequency of \textit{f} = 44.32 MHz by sweeping the magnetic field as shown in Fig.~\ref{11BNMRspectrum5-295K}. The FWHM increases with decreasing temperature and shows nearly constant($\sim$0.06 T) down to $\sim$50 K. Below 50 K, the FWHM slightly decreases, where the shape of the spectrum changes and the edges of the lines are smeared out. These results suggest a change in magnetic state around 50 K. Although it is not clear at present, it is interesting if the change relates to the strong enhancement of $\chi_b$ below 50 K as shown in Fig.~\ref{MPMSMTanddkidtinset}. NMR measurements on single crystals could provide additional information in this issue. This is the future work.

%The FWHM remains more or less constant down to 50 K. Further below 50 K, the FWHM decreases slightly accompanied by a change in the shape of the spectrum where the sharp edges become more rounded. Although there may be an anomalous feature in Fig.~\ref{TempvariationNMRparameter}(a) near 50 K (possibly associated with a spin reorientation transition as observed in switching in anisotropic magnetic susceptibility around 50 K as shown in Fig.~\ref{MPMSMTanddkidtinset})  consistent with previous literature~\cite{CHAI201552,POTASHNIKOV2019468}, this is hard to reconcile with a lack of any sharp features in \textit{M(T)} or \textit{R(T)} in this region. 

 %Interestingly, the coincidence of flipping of the magnitude of the antiferromagnetic susceptibility around 50 K as shown in fig.~\ref{MPMSMTanddkidtinset} and fluctuation in NMR order parameter around that temperature as shown in fig.~\ref{TempvariationNMRparameter} -  might be suggesting the spin reorientation transition from \textit{a}-\textit{c} plane to perpendicular to plane [010] direction. Since the susceptibility around room temperature is almost equal for [100] and [001]directions and the difference is more pounced below 50 K, it could be transition from room temperature 2 dimensional magnetic interaction to low temperature 3 dimensional character around 50 K similar to magnetism in R$_2$BaNiO, (R =Y,Er) oxides~\cite{Amador1990}.

%If we compare the $^{11}$B FWHM magnitude with the magnetic succeptibility, they are found to be proportional. This evidences that the $^{11}$B FWHM can serve as an order parameter.

Figure~\ref{TempvariationNMRparameter}(a) shows the temperature variation of the FWHM of the $^{11}$B NMR spectra between 5 - 457 K. Since the FWHM of the powder NMR spectrum in AFM state corresponds to the twice of the hyperfine field($H_{\rm hf}$) at the B site produced by Mn ordered moments, the temperature dependence of FWHM reflects the temperature dependence of the Mn sub-lattice magnetization. Therefore, one can obtain the critical exponent ($\beta$) of the order parameter using the formula $FWHM \propto (1 - \frac{T}{T_{\rm N}})^{\beta}$. The maximum value of $\beta = 0.21 \pm 0.02$ with $T_{\rm N}$ = 314 K was obtained by fitting the data points in the range 295 K - 315 K close to $T_{\rm N}$ as shown in Fig.~\ref{TempvariationNMRparameter}(b). Very nominal change was observed in the fitted $\beta$ parameter with the extension of fitted range toward the low temperature. The observed change in critical exponent $\beta$ was within the error bar for all the temperature range. This power law fittings of FWHM provided lower value of $\beta = 0.21 \pm 0.02$ (for 3D Heisenberg model $\beta \sim0.345$) suggesting a low dimensional magnetism as discussed in reference[\onlinecite{JMa2013}].

%identifying AlMn$_{2}$B$_{2}$ as an antiferromagnetic material with the transition temperature around 315 K. We fitted the observed temperature variation of FWHM order parameter  in the temperature range  295 - 315 K as $I = I_0 [1-(\frac{T}{T_{\rm N}})]^{\beta}$ with $T_{\rm N}$ = 314 K and $\beta = 0.21 \pm 0.02$ as shown in Fig.~\ref{TempvariationNMRparameter}(b).  

To study the dynamical properties of the Mn spins in high temperature range, the spin lattice relaxation rates ($\frac{1}{T_1}$) at the $^{11}$B site were measured from room temperature to 457 K. Figure~\ref{Temperaturevariationof1overT} shows the temperature dependence of $\frac{1}{T_1}$ where $\frac{1}{T_1}$ shows a clear peak at 315 K, evidencing again the AFM ordering. On the other hand, no clear anomaly in the temperature dependence of $\frac{1}{T_1}$ is observed around 390 K where the magnetic susceptibility exhibits a broad local maximum. Therefore, the broad maximum in the magnetic susceptibility is not associated with a magnetic ordering, but it could be attributed to a two dimensional magnetic character in AlMn$_2$B$_2$ as observed in 2D AFM compounds such as BaMN$_2$Si$_2$O$_7$~\cite{JMa2013}.

%Although there is, again, a clear signature of antiferromagnetic order near 315 K, no anomaly in ($\frac{1}{T_1}$) was observed around 390 K. The absence of anomaly in  relaxation rate ($T_1$) around 390 K combined with a broad hump in susceptibility data (see Fig.~\ref{CurieWeissOnly}(b)) at much higher temperature than antiferromagnetic transition  suggests a two dimensional magnetic character in AlMn$_{2}$B$_{2}$ as observed in  BaMn$_{2}$Si$_{2}$O$_{7}$~\cite{JMa2013}.

\section{Conclusions}
Structural, electrical transport and magnetic properties were studied on self flux grown single crystalline AlMn$_{2}$B$_{2}$ samples. All these measurements revealed that AlMn$_{2}$B$_{2}$ as an AFM compound with a transition temperature around 313$\pm$2 K. At higher temperature broad hump, well above the transition temperature, could be the signature of low dimensional magnetic interaction in AlMn$_{2}$B$_{2}$ above the room temperature. 

\section{Acknowledgement}
 Dr. Warren Straszheim is acknowledged for doing SEM on various samples. This research was supported by the Critical Materials Institute, an Energy Innovation Hub funded by the U.S. Department of Energy, Office of Energy Efficiency and Renewable Energy, Advanced Manufacturing Office. This work was also supported by the office of Basic Energy Sciences, Materials Sciences Division, U.S. DOE. This work was performed at the Ames Laboratory, operated for DOE by Iowa State University under Contract No. DE-AC02-07CH11358.\\*

\bibliographystyle{apsrev4-1} 
\bibliography{thebibliography}

\begin{thebibliography}{27}%
\makeatletter
\providecommand \@ifxundefined [1]{%
 \@ifx{#1\undefined}
}%
\providecommand \@ifnum [1]{%
 \ifnum #1\expandafter \@firstoftwo
 \else \expandafter \@secondoftwo
 \fi
}%
\providecommand \@ifx [1]{%
 \ifx #1\expandafter \@firstoftwo
 \else \expandafter \@secondoftwo
 \fi
}%
\providecommand \natexlab [1]{#1}%
\providecommand \enquote  [1]{``#1''}%
\providecommand \bibnamefont  [1]{#1}%
\providecommand \bibfnamefont [1]{#1}%
\providecommand \citenamefont [1]{#1}%
\providecommand \href@noop [0]{\@secondoftwo}%
\providecommand \href [0]{\begingroup \@sanitize@url \@href}%
\providecommand \@href[1]{\@@startlink{#1}\@@href}%
\providecommand \@@href[1]{\endgroup#1\@@endlink}%
\providecommand \@sanitize@url [0]{\catcode `\\12\catcode `\$12\catcode
  `\&12\catcode `\#12\catcode `\^12\catcode `\_12\catcode `\%12\relax}%
\providecommand \@@startlink[1]{}%
\providecommand \@@endlink[0]{}%
\providecommand \url  [0]{\begingroup\@sanitize@url \@url }%
\providecommand \@url [1]{\endgroup\@href {#1}{\urlprefix }}%
\providecommand \urlprefix  [0]{URL }%
\providecommand \Eprint [0]{\href }%
\providecommand \doibase [0]{http://dx.doi.org/}%
\providecommand \selectlanguage [0]{\@gobble}%
\providecommand \bibinfo  [0]{\@secondoftwo}%
\providecommand \bibfield  [0]{\@secondoftwo}%
\providecommand \translation [1]{[#1]}%
\providecommand \BibitemOpen [0]{}%
\providecommand \bibitemStop [0]{}%
\providecommand \bibitemNoStop [0]{.\EOS\space}%
\providecommand \EOS [0]{\spacefactor3000\relax}%
\providecommand \BibitemShut  [1]{\csname bibitem#1\endcsname}%
\let\auto@bib@innerbib\@empty
%</preamble>
\bibitem [{\citenamefont {Du}\ \emph {et~al.}(2015)\citenamefont {Du},
  \citenamefont {Chen}, \citenamefont {Yang}, \citenamefont {Wei},
  \citenamefont {Hua}, \citenamefont {Du}, \citenamefont {Wang}, \citenamefont
  {Liu}, \citenamefont {Han}, \citenamefont {Zhang},\ and\ \citenamefont
  {Yang}}]{Du2015}%
  \BibitemOpen
  \bibfield  {author} {\bibinfo {author} {\bibfnamefont {Qianheng}\
  \bibnamefont {Du}}, \bibinfo {author} {\bibfnamefont {Guofu}\ \bibnamefont
  {Chen}}, \bibinfo {author} {\bibfnamefont {Wenyun}\ \bibnamefont {Yang}},
  \bibinfo {author} {\bibfnamefont {Jianzhong}\ \bibnamefont {Wei}}, \bibinfo
  {author} {\bibfnamefont {Muxin}\ \bibnamefont {Hua}}, \bibinfo {author}
  {\bibfnamefont {Honglin}\ \bibnamefont {Du}}, \bibinfo {author}
  {\bibfnamefont {Changsheng}\ \bibnamefont {Wang}}, \bibinfo {author}
  {\bibfnamefont {Shunquan}\ \bibnamefont {Liu}}, \bibinfo {author}
  {\bibfnamefont {Jingzhi}\ \bibnamefont {Han}}, \bibinfo {author}
  {\bibfnamefont {Yan}\ \bibnamefont {Zhang}}, \ and\ \bibinfo {author}
  {\bibfnamefont {Jinbo}\ \bibnamefont {Yang}},\ }\bibfield  {title} {\enquote
  {\bibinfo {title} {{Magnetic frustration and magnetocaloric effect in
  AlFe$_{2-x}$Mn$_x$B$_2$(x= 0{\textendash}0.5) ribbons}},}\ }\href {\doibase
  10.1088/0022-3727/48/33/335001} {\bibfield  {journal} {\bibinfo  {journal}
  {Journal of Physics D: Applied Physics}\ }\textbf {\bibinfo {volume} {48}},\
  \bibinfo {pages} {335001} (\bibinfo {year} {2015})}\BibitemShut {NoStop}%
\bibitem [{\citenamefont {Tan}\ \emph {et~al.}(2013)\citenamefont {Tan},
  \citenamefont {Chai}, \citenamefont {Thompson},\ and\ \citenamefont
  {Shatruk}}]{Tan2013}%
  \BibitemOpen
  \bibfield  {author} {\bibinfo {author} {\bibfnamefont {Xiaoyan}\ \bibnamefont
  {Tan}}, \bibinfo {author} {\bibfnamefont {Ping}\ \bibnamefont {Chai}},
  \bibinfo {author} {\bibfnamefont {Corey~M.}\ \bibnamefont {Thompson}}, \ and\
  \bibinfo {author} {\bibfnamefont {Michael}\ \bibnamefont {Shatruk}},\
  }\bibfield  {title} {\enquote {\bibinfo {title} {{Magnetocaloric Effect in
  AlFe$_2$B$_2$: Toward Magnetic Refrigerants from Earth-Abundant Elements}},}\
  }\href {\doibase 10.1021/ja404107p} {\bibfield  {journal} {\bibinfo
  {journal} {Journal of the American Chemical Society}\ }\textbf {\bibinfo
  {volume} {135}},\ \bibinfo {pages} {9553--9557} (\bibinfo {year} {2013})},\
  \bibinfo {note} {pMID: 23731263},\ \Eprint
  {http://arxiv.org/abs/https://doi.org/10.1021/ja404107p}
  {https://doi.org/10.1021/ja404107p} \BibitemShut {NoStop}%
\bibitem [{\citenamefont {Lamichhane}\ \emph {et~al.}(2018)\citenamefont
  {Lamichhane}, \citenamefont {Xiang}, \citenamefont {Lin}, \citenamefont
  {Pandey}, \citenamefont {Parker}, \citenamefont {Kim}, \citenamefont {Zhou},
  \citenamefont {Kramer}, \citenamefont {Bud'ko},\ and\ \citenamefont
  {Canfield}}]{LamichhaneAlFeB2018}%
  \BibitemOpen
  \bibfield  {author} {\bibinfo {author} {\bibfnamefont {Tej~N.}\ \bibnamefont
  {Lamichhane}}, \bibinfo {author} {\bibfnamefont {Li}~\bibnamefont {Xiang}},
  \bibinfo {author} {\bibfnamefont {Qisheng}\ \bibnamefont {Lin}}, \bibinfo
  {author} {\bibfnamefont {Tribhuwan}\ \bibnamefont {Pandey}}, \bibinfo
  {author} {\bibfnamefont {David~S.}\ \bibnamefont {Parker}}, \bibinfo {author}
  {\bibfnamefont {Tae-Hoon}\ \bibnamefont {Kim}}, \bibinfo {author}
  {\bibfnamefont {Lin}\ \bibnamefont {Zhou}}, \bibinfo {author} {\bibfnamefont
  {Matthew~J.}\ \bibnamefont {Kramer}}, \bibinfo {author} {\bibfnamefont
  {Sergey~L.}\ \bibnamefont {Bud'ko}}, \ and\ \bibinfo {author} {\bibfnamefont
  {Paul~C.}\ \bibnamefont {Canfield}},\ }\bibfield  {title} {\enquote {\bibinfo
  {title} {{Magnetic properties of single crystalline itinerant ferromagnet
  ${\mathrm{AlFe}}_{2}{\mathrm{B}}_{2}$}},}\ }\href {\doibase
  10.1103/PhysRevMaterials.2.084408} {\bibfield  {journal} {\bibinfo  {journal}
  {Phys. Rev. Materials}\ }\textbf {\bibinfo {volume} {2}},\ \bibinfo {pages}
  {084408} (\bibinfo {year} {2018})}\BibitemShut {NoStop}%
\bibitem [{\citenamefont {Barua}\ \emph {et~al.}(2018)\citenamefont {Barua},
  \citenamefont {Lejeune}, \citenamefont {Ke}, \citenamefont {Hadjipanayis},
  \citenamefont {Levin}, \citenamefont {McCallum}, \citenamefont {Kramer},\
  and\ \citenamefont {Lewis}}]{BARUA2018505}%
  \BibitemOpen
  \bibfield  {author} {\bibinfo {author} {\bibfnamefont {R.}~\bibnamefont
  {Barua}}, \bibinfo {author} {\bibfnamefont {B.T.}\ \bibnamefont {Lejeune}},
  \bibinfo {author} {\bibfnamefont {L.}~\bibnamefont {Ke}}, \bibinfo {author}
  {\bibfnamefont {G.}~\bibnamefont {Hadjipanayis}}, \bibinfo {author}
  {\bibfnamefont {E.M.}\ \bibnamefont {Levin}}, \bibinfo {author}
  {\bibfnamefont {R.W.}\ \bibnamefont {McCallum}}, \bibinfo {author}
  {\bibfnamefont {M.J.}\ \bibnamefont {Kramer}}, \ and\ \bibinfo {author}
  {\bibfnamefont {L.H.}\ \bibnamefont {Lewis}},\ }\bibfield  {title} {\enquote
  {\bibinfo {title} {{Anisotropic magnetocaloric response in AlFe$_2$B$_2$}},}\
  }\href {\doibase 10.1016/j.jallcom.2018.02.205} {\bibfield  {journal}
  {\bibinfo  {journal} {Journal of Alloys and Compounds}\ }\textbf {\bibinfo
  {volume} {745}},\ \bibinfo {pages} {505 -- 512} (\bibinfo {year}
  {2018})}\BibitemShut {NoStop}%
\bibitem [{\citenamefont {Chai}\ \emph {et~al.}(2015)\citenamefont {Chai},
  \citenamefont {Stoian}, \citenamefont {Tan}, \citenamefont {Dube},\ and\
  \citenamefont {Shatruk}}]{CHAI201552}%
  \BibitemOpen
  \bibfield  {author} {\bibinfo {author} {\bibfnamefont {Ping}\ \bibnamefont
  {Chai}}, \bibinfo {author} {\bibfnamefont {Sebastian~A.}\ \bibnamefont
  {Stoian}}, \bibinfo {author} {\bibfnamefont {Xiaoyan}\ \bibnamefont {Tan}},
  \bibinfo {author} {\bibfnamefont {Paul~A.}\ \bibnamefont {Dube}}, \ and\
  \bibinfo {author} {\bibfnamefont {Michael}\ \bibnamefont {Shatruk}},\
  }\bibfield  {title} {\enquote {\bibinfo {title} {{Investigation of magnetic
  properties and electronic structure of layered-structure borides AlT$_2$B$_2$
  (T=Fe, Mn, Cr) and AlFe$_{2-x}$Mn$_x$B$_2$}},}\ }\href {\doibase
  10.1016/j.jssc.2014.04.027} {\bibfield  {journal} {\bibinfo  {journal}
  {Journal of Solid State Chemistry}\ }\textbf {\bibinfo {volume} {224}},\
  \bibinfo {pages} {52 -- 61} (\bibinfo {year} {2015})}\BibitemShut {NoStop}%
\bibitem [{\citenamefont {Ke}\ \emph {et~al.}(2017)\citenamefont {Ke},
  \citenamefont {Harmon},\ and\ \citenamefont {Kramer}}]{KeLiqin2017}%
  \BibitemOpen
  \bibfield  {author} {\bibinfo {author} {\bibfnamefont {Liqin}\ \bibnamefont
  {Ke}}, \bibinfo {author} {\bibfnamefont {Bruce~N.}\ \bibnamefont {Harmon}}, \
  and\ \bibinfo {author} {\bibfnamefont {Matthew~J.}\ \bibnamefont {Kramer}},\
  }\bibfield  {title} {\enquote {\bibinfo {title} {{Electronic structure and
  magnetic properties in ${T}_{2}{\mathrm{AlB}}_{2}$ ($T$=Fe, Mn, Cr, Co, and
  Ni) and their alloys}},}\ }\href {\doibase 10.1103/PhysRevB.95.104427}
  {\bibfield  {journal} {\bibinfo  {journal} {Phys. Rev. B}\ }\textbf {\bibinfo
  {volume} {95}},\ \bibinfo {pages} {104427} (\bibinfo {year}
  {2017})}\BibitemShut {NoStop}%
\bibitem [{\citenamefont {Potashnikov}\ \emph {et~al.}(2019)\citenamefont
  {Potashnikov}, \citenamefont {Caspi}, \citenamefont {Pesach}, \citenamefont
  {Hoser}, \citenamefont {Kota}, \citenamefont {Verger}, \citenamefont
  {Barsoum}, \citenamefont {Felner}, \citenamefont {Keren},\ and\ \citenamefont
  {Rivin}}]{POTASHNIKOV2019468}%
  \BibitemOpen
  \bibfield  {author} {\bibinfo {author} {\bibfnamefont {D.}~\bibnamefont
  {Potashnikov}}, \bibinfo {author} {\bibfnamefont {E.N.}\ \bibnamefont
  {Caspi}}, \bibinfo {author} {\bibfnamefont {A.}~\bibnamefont {Pesach}},
  \bibinfo {author} {\bibfnamefont {A.}~\bibnamefont {Hoser}}, \bibinfo
  {author} {\bibfnamefont {S.}~\bibnamefont {Kota}}, \bibinfo {author}
  {\bibfnamefont {L.}~\bibnamefont {Verger}}, \bibinfo {author} {\bibfnamefont
  {M.W.}\ \bibnamefont {Barsoum}}, \bibinfo {author} {\bibfnamefont
  {I.}~\bibnamefont {Felner}}, \bibinfo {author} {\bibfnamefont
  {A.}~\bibnamefont {Keren}}, \ and\ \bibinfo {author} {\bibfnamefont
  {O.}~\bibnamefont {Rivin}},\ }\bibfield  {title} {\enquote {\bibinfo {title}
  {{Magnetic ordering in the nano-laminar ternary Mn$_2$AlB$_2 $using neutron
  and X-ray diffraction}},}\ }\href {\doibase 10.1016/j.jmmm.2018.09.078}
  {\bibfield  {journal} {\bibinfo  {journal} {Journal of Magnetism and Magnetic
  Materials}\ }\textbf {\bibinfo {volume} {471}},\ \bibinfo {pages} {468 --
  474} (\bibinfo {year} {2019})}\BibitemShut {NoStop}%
\bibitem [{\citenamefont {Verger}\ \emph {et~al.}(2018)\citenamefont {Verger},
  \citenamefont {Kota}, \citenamefont {Roussel}, \citenamefont {Ouisse},\ and\
  \citenamefont {Barsoum}}]{Verger2018}%
  \BibitemOpen
  \bibfield  {author} {\bibinfo {author} {\bibfnamefont {L.}~\bibnamefont
  {Verger}}, \bibinfo {author} {\bibfnamefont {S.}~\bibnamefont {Kota}},
  \bibinfo {author} {\bibfnamefont {H.}~\bibnamefont {Roussel}}, \bibinfo
  {author} {\bibfnamefont {T.}~\bibnamefont {Ouisse}}, \ and\ \bibinfo {author}
  {\bibfnamefont {M.~W.}\ \bibnamefont {Barsoum}},\ }\bibfield  {title}
  {\enquote {\bibinfo {title} {{Anisotropic thermal expansions of select
  layered ternary transition metal borides: MoAlB, Cr$_2$AlB$_2$,
  Mn$_2$AlB$_2$, and Fe$_2$AlB$_2$}},}\ }\href {\doibase 10.1063/1.5054379}
  {\bibfield  {journal} {\bibinfo  {journal} {Journal of Applied Physics}\
  }\textbf {\bibinfo {volume} {124}},\ \bibinfo {pages} {205108} (\bibinfo
  {year} {2018})}\BibitemShut {NoStop}%
\bibitem [{\citenamefont {Belashchenko}\ \emph {et~al.}(2015)\citenamefont
  {Belashchenko}, \citenamefont {Ke}, \citenamefont {Dane}, \citenamefont
  {Benedict}, \citenamefont {Lamichhane}, \citenamefont {Taufour},
  \citenamefont {Jesche}, \citenamefont {Bud'ko}, \citenamefont {Canfield},\
  and\ \citenamefont {Antropov}}]{Belashchenko2015}%
  \BibitemOpen
  \bibfield  {author} {\bibinfo {author} {\bibfnamefont {Kirill~D.}\
  \bibnamefont {Belashchenko}}, \bibinfo {author} {\bibfnamefont {Liqin}\
  \bibnamefont {Ke}}, \bibinfo {author} {\bibfnamefont {Markus}\ \bibnamefont
  {Dane}}, \bibinfo {author} {\bibfnamefont {Lorin~X.}\ \bibnamefont
  {Benedict}}, \bibinfo {author} {\bibfnamefont {Tej~Nath}\ \bibnamefont
  {Lamichhane}}, \bibinfo {author} {\bibfnamefont {Valentin}\ \bibnamefont
  {Taufour}}, \bibinfo {author} {\bibfnamefont {Anton}\ \bibnamefont {Jesche}},
  \bibinfo {author} {\bibfnamefont {Sergey~L.}\ \bibnamefont {Bud'ko}},
  \bibinfo {author} {\bibfnamefont {Paul~C.}\ \bibnamefont {Canfield}}, \ and\
  \bibinfo {author} {\bibfnamefont {Vladimir~P.}\ \bibnamefont {Antropov}},\
  }\bibfield  {title} {\enquote {\bibinfo {title} {{Origin of the spin
  reorientation transitions in (Fe$_{1 - x}$Co$_x$)$_2$B alloys}},}\ }\href
  {\doibase 10.1063/1.4908056} {\bibfield  {journal} {\bibinfo  {journal}
  {Applied Physics Letters}\ }\textbf {\bibinfo {volume} {106}},\ \bibinfo
  {pages} {062408} (\bibinfo {year} {2015})},\ \Eprint
  {http://arxiv.org/abs/https://doi.org/10.1063/1.4908056}
  {https://doi.org/10.1063/1.4908056} \BibitemShut {NoStop}%
\bibitem [{\citenamefont {Canfield}\ and\ \citenamefont
  {Fisher}(2001)}]{Canfield2001}%
  \BibitemOpen
  \bibfield  {author} {\bibinfo {author} {\bibfnamefont {Paul~C.}\ \bibnamefont
  {Canfield}}\ and\ \bibinfo {author} {\bibfnamefont {Ian~R.}\ \bibnamefont
  {Fisher}},\ }\bibfield  {title} {\enquote {\bibinfo {title} {High-temperature
  solution growth of intermetallic single crystals and quasicrystals},}\ }\href
  {\doibase 10.1016/S0022-0248(01)00827-2} {\bibfield  {journal} {\bibinfo
  {journal} {Journal of Crystal Growth}\ ,\ \bibinfo {pages} {155 -- 161}}
  (\bibinfo {year} {2001})}\BibitemShut {NoStop}%
\bibitem [{\citenamefont {Meier}\ \emph {et~al.}(2017)\citenamefont {Meier},
  \citenamefont {Kong}, \citenamefont {Bud'ko},\ and\ \citenamefont
  {Canfield}}]{Meier2017}%
  \BibitemOpen
  \bibfield  {author} {\bibinfo {author} {\bibfnamefont {W.~R.}\ \bibnamefont
  {Meier}}, \bibinfo {author} {\bibfnamefont {T.}~\bibnamefont {Kong}},
  \bibinfo {author} {\bibfnamefont {S.~L.}\ \bibnamefont {Bud'ko}}, \ and\
  \bibinfo {author} {\bibfnamefont {P.~C.}\ \bibnamefont {Canfield}},\
  }\bibfield  {title} {\enquote {\bibinfo {title} {{Optimization of the crystal
  growth of the superconductor CaKFe$_4$As$_4$ from solution in the FeAS -
  CaFe$_2$As$_2$ - KFe$_2$As$_2$ system}},}\ }\href {\doibase
  10.1103/PhysRevMaterials.1.013401} {\bibfield  {journal} {\bibinfo  {journal}
  {Phys. Rev. Materials}\ }\textbf {\bibinfo {volume} {1}},\ \bibinfo {pages}
  {013401} (\bibinfo {year} {2017})}\BibitemShut {NoStop}%
\bibitem [{\citenamefont {Canfield}\ \emph {et~al.}(2016)\citenamefont
  {Canfield}, \citenamefont {Kong}, \citenamefont {Kaluarachchi},\ and\
  \citenamefont {Jo}}]{Canfieldfrittedcrucible2016}%
  \BibitemOpen
  \bibfield  {author} {\bibinfo {author} {\bibfnamefont {Paul~C.}\ \bibnamefont
  {Canfield}}, \bibinfo {author} {\bibfnamefont {Tai}\ \bibnamefont {Kong}},
  \bibinfo {author} {\bibfnamefont {Udhara~S.}\ \bibnamefont {Kaluarachchi}}, \
  and\ \bibinfo {author} {\bibfnamefont {Na~Hyun}\ \bibnamefont {Jo}},\
  }\bibfield  {title} {\enquote {\bibinfo {title} {Use of frit-disc crucibles
  for routine and exploratory solution growth of single crystalline samples},}\
  }\href {\doibase 10.1080/14786435.2015.1122248} {\bibfield  {journal}
  {\bibinfo  {journal} {Philosophical Magazine}\ }\textbf {\bibinfo {volume}
  {96}},\ \bibinfo {pages} {84 -- 92} (\bibinfo {year} {2016})}\BibitemShut
  {NoStop}%
\bibitem [{\citenamefont {Lamichhane}\ \emph {et~al.}(2016)\citenamefont
  {Lamichhane}, \citenamefont {Taufour}, \citenamefont {Masters}, \citenamefont
  {Parker}, \citenamefont {Kaluarachchi}, \citenamefont {Thimmaiah},
  \citenamefont {Bud'ko},\ and\ \citenamefont {Canfield}}]{LamichhaneZrMnP}%
  \BibitemOpen
  \bibfield  {author} {\bibinfo {author} {\bibfnamefont {Tej~N.}\ \bibnamefont
  {Lamichhane}}, \bibinfo {author} {\bibfnamefont {Valentin}\ \bibnamefont
  {Taufour}}, \bibinfo {author} {\bibfnamefont {Morgan~W.}\ \bibnamefont
  {Masters}}, \bibinfo {author} {\bibfnamefont {David~S.}\ \bibnamefont
  {Parker}}, \bibinfo {author} {\bibfnamefont {Udhara~S.}\ \bibnamefont
  {Kaluarachchi}}, \bibinfo {author} {\bibfnamefont {Srinivasa}\ \bibnamefont
  {Thimmaiah}}, \bibinfo {author} {\bibfnamefont {Sergey~L.}\ \bibnamefont
  {Bud'ko}}, \ and\ \bibinfo {author} {\bibfnamefont {Paul~C.}\ \bibnamefont
  {Canfield}},\ }\bibfield  {title} {\enquote {\bibinfo {title} {{Discovery of
  ferromagnetism with large magnetic anisotropy in ZrMnP and HfMnP}},}\ }\href
  {\doibase 10.1063/1.4961933} {\bibfield  {journal} {\bibinfo  {journal}
  {Applied Physics Letters}\ }\textbf {\bibinfo {volume} {109}},\ \bibinfo
  {pages} {092402} (\bibinfo {year} {2016})}\BibitemShut {NoStop}%
\bibitem [{\citenamefont {Larson}\ and\ \citenamefont
  {Dreele}(2004)}]{Larson2004}%
  \BibitemOpen
  \bibfield  {author} {\bibinfo {author} {\bibfnamefont {A.~C.}\ \bibnamefont
  {Larson}}\ and\ \bibinfo {author} {\bibfnamefont {R.~B.~Von}\ \bibnamefont
  {Dreele}},\ }\bibfield  {title} {\enquote {\bibinfo {title} {General
  structure analysis system},}\ }\href@noop {} {\bibfield  {journal} {\bibinfo
  {journal} {Los Alamos National Laboratory Report No. LAUR 86-748}\ }
  (\bibinfo {year} {2004})}\BibitemShut {NoStop}%
\bibitem [{\citenamefont {Toby}(2001)}]{Toby2001JAC}%
  \BibitemOpen
  \bibfield  {author} {\bibinfo {author} {\bibfnamefont {Brian~H.}\
  \bibnamefont {Toby}},\ }\bibfield  {title} {\enquote {\bibinfo {title} {{{\it
  EXPGUI}, a graphical user interface for {\it GSAS}}},}\ }\href {\doibase
  10.1107/S0021889801002242} {\bibfield  {journal} {\bibinfo  {journal}
  {Journal of Applied Crystallography}\ }\textbf {\bibinfo {volume} {34}},\
  \bibinfo {pages} {210--213} (\bibinfo {year} {2001})}\BibitemShut {NoStop}%
\bibitem [{\citenamefont {Jesche}\ \emph {et~al.}(2016)\citenamefont {Jesche},
  \citenamefont {Fix}, \citenamefont {Kreyssig}, \citenamefont {Meier},\ and\
  \citenamefont {Canfield}}]{AJesche2016}%
  \BibitemOpen
  \bibfield  {author} {\bibinfo {author} {\bibfnamefont {A.}~\bibnamefont
  {Jesche}}, \bibinfo {author} {\bibfnamefont {M.}~\bibnamefont {Fix}},
  \bibinfo {author} {\bibfnamefont {A.}~\bibnamefont {Kreyssig}}, \bibinfo
  {author} {\bibfnamefont {W.~R.}\ \bibnamefont {Meier}}, \ and\ \bibinfo
  {author} {\bibfnamefont {P.~C.}\ \bibnamefont {Canfield}},\ }\bibfield
  {title} {\enquote {\bibinfo {title} {X-ray diffraction on large single
  crystals using a powder diffractometer},}\ }\href {\doibase
  10.1080/14786435.2016.1192725} {\bibfield  {journal} {\bibinfo  {journal}
  {Philosophical Magazine}\ }\textbf {\bibinfo {volume} {96}},\ \bibinfo
  {pages} {2115--2124} (\bibinfo {year} {2016})},\ \Eprint
  {http://arxiv.org/abs/https://doi.org/10.1080/14786435.2016.1192725}
  {https://doi.org/10.1080/14786435.2016.1192725} \BibitemShut {NoStop}%
\bibitem [{\citenamefont {Fisher}\ and\ \citenamefont
  {Langer}(1968)}]{MFisher1968}%
  \BibitemOpen
  \bibfield  {author} {\bibinfo {author} {\bibfnamefont {Michael~E.}\
  \bibnamefont {Fisher}}\ and\ \bibinfo {author} {\bibfnamefont {J.~S.}\
  \bibnamefont {Langer}},\ }\bibfield  {title} {\enquote {\bibinfo {title}
  {Resistive anomalies at magnetic critical points},}\ }\href {\doibase
  10.1103/PhysRevLett.20.665} {\bibfield  {journal} {\bibinfo  {journal} {Phys.
  Rev. Lett.}\ }\textbf {\bibinfo {volume} {20}},\ \bibinfo {pages} {665--668}
  (\bibinfo {year} {1968})}\BibitemShut {NoStop}%
\bibitem [{\citenamefont {Fisher}(1962)}]{MFishercriterion}%
  \BibitemOpen
  \bibfield  {author} {\bibinfo {author} {\bibfnamefont {Michael~E.}\
  \bibnamefont {Fisher}},\ }\bibfield  {title} {\enquote {\bibinfo {title}
  {Relation between the specific heat and susceptibility of an
  antiferromagnet},}\ }\href {\doibase 10.1080/14786436208213705} {\bibfield
  {journal} {\bibinfo  {journal} {The Philosophical Magazine: A Journal of
  Theoretical Experimental and Applied Physics}\ }\textbf {\bibinfo {volume}
  {7}},\ \bibinfo {pages} {1731--1743} (\bibinfo {year} {1962})},\ \Eprint
  {http://arxiv.org/abs/https://doi.org/10.1080/14786436208213705}
  {https://doi.org/10.1080/14786436208213705} \BibitemShut {NoStop}%
\bibitem [{\citenamefont {Ma}\ \emph {et~al.}(2013)\citenamefont {Ma},
  \citenamefont {Dela~Cruz}, \citenamefont {Hong}, \citenamefont {Tian},
  \citenamefont {Aczel}, \citenamefont {Chi}, \citenamefont {Yan},
  \citenamefont {Dun}, \citenamefont {Zhou},\ and\ \citenamefont
  {Matsuda}}]{JMa2013}%
  \BibitemOpen
  \bibfield  {author} {\bibinfo {author} {\bibfnamefont {J.}~\bibnamefont
  {Ma}}, \bibinfo {author} {\bibfnamefont {C.~D.}\ \bibnamefont {Dela~Cruz}},
  \bibinfo {author} {\bibfnamefont {Tao}\ \bibnamefont {Hong}}, \bibinfo
  {author} {\bibfnamefont {W.}~\bibnamefont {Tian}}, \bibinfo {author}
  {\bibfnamefont {A.~A.}\ \bibnamefont {Aczel}}, \bibinfo {author}
  {\bibfnamefont {Songxue}\ \bibnamefont {Chi}}, \bibinfo {author}
  {\bibfnamefont {J.-Q.}\ \bibnamefont {Yan}}, \bibinfo {author} {\bibfnamefont
  {Z.~L.}\ \bibnamefont {Dun}}, \bibinfo {author} {\bibfnamefont {H.~D.}\
  \bibnamefont {Zhou}}, \ and\ \bibinfo {author} {\bibfnamefont
  {M.}~\bibnamefont {Matsuda}},\ }\bibfield  {title} {\enquote {\bibinfo
  {title} {{Magnetic phase transition in the low-dimensional compound
  BaMn$_{2}$Si$_{2}$O$_{7}$}},}\ }\href {\doibase 10.1103/PhysRevB.88.144405}
  {\bibfield  {journal} {\bibinfo  {journal} {Phys. Rev. B}\ }\textbf {\bibinfo
  {volume} {88}},\ \bibinfo {pages} {144405} (\bibinfo {year}
  {2013})}\BibitemShut {NoStop}%
\bibitem [{\citenamefont {Vasil'ev}\ \emph {et~al.}(2000)\citenamefont
  {Vasil'ev}, \citenamefont {Ponomarenko}, \citenamefont {Manaka},
  \citenamefont {Yamada}, \citenamefont {Isobe},\ and\ \citenamefont
  {Ueda}}]{VASILEV20001619}%
  \BibitemOpen
  \bibfield  {author} {\bibinfo {author} {\bibfnamefont {A.N}\ \bibnamefont
  {Vasil'ev}}, \bibinfo {author} {\bibfnamefont {L.A}\ \bibnamefont
  {Ponomarenko}}, \bibinfo {author} {\bibfnamefont {H}~\bibnamefont {Manaka}},
  \bibinfo {author} {\bibfnamefont {I}~\bibnamefont {Yamada}}, \bibinfo
  {author} {\bibfnamefont {M}~\bibnamefont {Isobe}}, \ and\ \bibinfo {author}
  {\bibfnamefont {Y}~\bibnamefont {Ueda}},\ }\bibfield  {title} {\enquote
  {\bibinfo {title} {{Quasi-one-dimensional antiferromagnetic spinel compound
  LiCuVO$_4$}},}\ }\href {\doibase
  https://doi.org/10.1016/S0921-4526(99)02739-8} {\bibfield  {journal}
  {\bibinfo  {journal} {Physica B: Condensed Matter}\ }\textbf {\bibinfo
  {volume} {284-288}},\ \bibinfo {pages} {1619 -- 1620} (\bibinfo {year}
  {2000})}\BibitemShut {NoStop}%
\bibitem [{\citenamefont {Kim}\ \emph {et~al.}(1998)\citenamefont {Kim},
  \citenamefont {Greven}, \citenamefont {Wiese},\ and\ \citenamefont
  {Birgeneau}}]{Kim1998}%
  \BibitemOpen
  \bibfield  {author} {\bibinfo {author} {\bibfnamefont {Y.J.}\ \bibnamefont
  {Kim}}, \bibinfo {author} {\bibfnamefont {M.}~\bibnamefont {Greven}},
  \bibinfo {author} {\bibfnamefont {U.-J.}\ \bibnamefont {Wiese}}, \ and\
  \bibinfo {author} {\bibfnamefont {R.J.}\ \bibnamefont {Birgeneau}},\
  }\bibfield  {title} {\enquote {\bibinfo {title} {Monte-carlo study of
  correlations in quantum spin chains at non-zero temperature},}\ }\href
  {\doibase 10.1007/s100510050382} {\bibfield  {journal} {\bibinfo  {journal}
  {The European Physical Journal B - Condensed Matter and Complex Systems}\
  }\textbf {\bibinfo {volume} {4}},\ \bibinfo {pages} {291--297} (\bibinfo
  {year} {1998})}\BibitemShut {NoStop}%
\bibitem [{\citenamefont {Bonner}\ and\ \citenamefont
  {Fisher}(1964)}]{Bonner1964}%
  \BibitemOpen
  \bibfield  {author} {\bibinfo {author} {\bibfnamefont {Jill~C.}\ \bibnamefont
  {Bonner}}\ and\ \bibinfo {author} {\bibfnamefont {Michael~E.}\ \bibnamefont
  {Fisher}},\ }\bibfield  {title} {\enquote {\bibinfo {title} {Linear magnetic
  chains with anisotropic coupling},}\ }\href@noop {} {\bibfield  {journal}
  {\bibinfo  {journal} {Physical Review}\ }\textbf {\bibinfo {volume} {135}},\
  \bibinfo {pages} {640 --658} (\bibinfo {year} {1964})}\BibitemShut {NoStop}%
\bibitem [{\citenamefont {Dingle}\ \emph {et~al.}(1969)\citenamefont {Dingle},
  \citenamefont {Lines},\ and\ \citenamefont {Holt}}]{Dingle1969}%
  \BibitemOpen
  \bibfield  {author} {\bibinfo {author} {\bibfnamefont {R.}~\bibnamefont
  {Dingle}}, \bibinfo {author} {\bibfnamefont {M.~E.}\ \bibnamefont {Lines}}, \
  and\ \bibinfo {author} {\bibfnamefont {S.~L.}\ \bibnamefont {Holt}},\
  }\bibfield  {title} {\enquote {\bibinfo {title} {{Linear-Chain
  Antiferromagnetism in [${(\mathrm{C}{\mathrm{H}}_{3})}_{4}$N]
  [Mn${\mathrm{Cl}}_{3}$]}},}\ }\href {\doibase 10.1103/PhysRev.187.643}
  {\bibfield  {journal} {\bibinfo  {journal} {Phys. Rev.}\ }\textbf {\bibinfo
  {volume} {187}},\ \bibinfo {pages} {643--648} (\bibinfo {year}
  {1969})}\BibitemShut {NoStop}%
\bibitem [{\citenamefont {Nath}\ \emph {et~al.}(2014)\citenamefont {Nath},
  \citenamefont {Ranjith}, \citenamefont {Roy}, \citenamefont {Johnston},
  \citenamefont {Furukawa},\ and\ \citenamefont {Tsirlin}}]{Nath2014}%
  \BibitemOpen
  \bibfield  {author} {\bibinfo {author} {\bibfnamefont {R.}~\bibnamefont
  {Nath}}, \bibinfo {author} {\bibfnamefont {K.~M.}\ \bibnamefont {Ranjith}},
  \bibinfo {author} {\bibfnamefont {B.}~\bibnamefont {Roy}}, \bibinfo {author}
  {\bibfnamefont {D.~C.}\ \bibnamefont {Johnston}}, \bibinfo {author}
  {\bibfnamefont {Y.}~\bibnamefont {Furukawa}}, \ and\ \bibinfo {author}
  {\bibfnamefont {A.~A.}\ \bibnamefont {Tsirlin}},\ }\bibfield  {title}
  {\enquote {\bibinfo {title} {{Magnetic transitions in the spin-$\frac{5}{2}$
  frustrated magnet $\mathrm{Bi}{\mathrm{Mn}}_{2}{\mathrm{PO}}_{6}$ and strong
  lattice softening in $\mathrm{Bi}{\mathrm{Mn}}_{2}{\mathrm{PO}}_{6}$ and
  $\mathrm{Bi}{\mathrm{Zn}}_{2}{\mathrm{PO}}_{6}$ below 200 K}},}\ }\href
  {\doibase 10.1103/PhysRevB.90.024431} {\bibfield  {journal} {\bibinfo
  {journal} {Phys. Rev. B}\ }\textbf {\bibinfo {volume} {90}},\ \bibinfo
  {pages} {024431} (\bibinfo {year} {2014})}\BibitemShut {NoStop}%
\bibitem [{\citenamefont {Pedrini}\ \emph {et~al.}(2004)\citenamefont
  {Pedrini}, \citenamefont {Gavilano}, \citenamefont {Rau}, \citenamefont
  {Ott}, \citenamefont {Kazakov}, \citenamefont {Karpinski},\ and\
  \citenamefont {Wessel}}]{Pedrini2004}%
  \BibitemOpen
  \bibfield  {author} {\bibinfo {author} {\bibfnamefont {B.}~\bibnamefont
  {Pedrini}}, \bibinfo {author} {\bibfnamefont {J.~L.}\ \bibnamefont
  {Gavilano}}, \bibinfo {author} {\bibfnamefont {D.}~\bibnamefont {Rau}},
  \bibinfo {author} {\bibfnamefont {H.~R.}\ \bibnamefont {Ott}}, \bibinfo
  {author} {\bibfnamefont {S.~M.}\ \bibnamefont {Kazakov}}, \bibinfo {author}
  {\bibfnamefont {J.}~\bibnamefont {Karpinski}}, \ and\ \bibinfo {author}
  {\bibfnamefont {S.}~\bibnamefont {Wessel}},\ }\bibfield  {title} {\enquote
  {\bibinfo {title} {{NMR and dc susceptibility studies of
  ${\mathrm{NaVGe}}_{2}{\mathrm{O}}_{6}$}},}\ }\href {\doibase
  10.1103/PhysRevB.70.024421} {\bibfield  {journal} {\bibinfo  {journal} {Phys.
  Rev. B}\ }\textbf {\bibinfo {volume} {70}},\ \bibinfo {pages} {024421}
  (\bibinfo {year} {2004})}\BibitemShut {NoStop}%
\bibitem [{\citenamefont {Kikuchi}\ \emph {et~al.}(2000)\citenamefont
  {Kikuchi}, \citenamefont {Ishiguchi}, \citenamefont {Motoya}, \citenamefont
  {Itoh}, \citenamefont {Inari}, \citenamefont {Eguchi},\ and\ \citenamefont
  {Akimitsu}}]{Kikuchi2000}%
  \BibitemOpen
  \bibfield  {author} {\bibinfo {author} {\bibfnamefont {Jun}\ \bibnamefont
  {Kikuchi}}, \bibinfo {author} {\bibfnamefont {Kazuhiro}\ \bibnamefont
  {Ishiguchi}}, \bibinfo {author} {\bibfnamefont {Kiyoichiro}\ \bibnamefont
  {Motoya}}, \bibinfo {author} {\bibfnamefont {Masayuki}\ \bibnamefont {Itoh}},
  \bibinfo {author} {\bibfnamefont {Kazunori}\ \bibnamefont {Inari}}, \bibinfo
  {author} {\bibfnamefont {Naotoshi}\ \bibnamefont {Eguchi}}, \ and\ \bibinfo
  {author} {\bibfnamefont {Jun}\ \bibnamefont {Akimitsu}},\ }\bibfield  {title}
  {\enquote {\bibinfo {title} {{NMR and Neutron Scattering Studies of Quasi
  One-Dimensional Magnet CuV$_2$O$_6$}},}\ }\href {\doibase
  10.1143/JPSJ.69.2660} {\bibfield  {journal} {\bibinfo  {journal} {Journal of
  the Physical Society of Japan}\ }\textbf {\bibinfo {volume} {69}},\ \bibinfo
  {pages} {2660--2668} (\bibinfo {year} {2000})}\BibitemShut {NoStop}%
\bibitem [{\citenamefont {Ideta}\ \emph {et~al.}(2012)\citenamefont {Ideta},
  \citenamefont {Kawasaki}, \citenamefont {Kishimoto}, \citenamefont {Ohno},
  \citenamefont {Michihiro}, \citenamefont {He}, \citenamefont {Ueda},\ and\
  \citenamefont {Itoh}}]{Ideta2012}%
  \BibitemOpen
  \bibfield  {author} {\bibinfo {author} {\bibfnamefont {Yukiichi}\
  \bibnamefont {Ideta}}, \bibinfo {author} {\bibfnamefont {Yu}~\bibnamefont
  {Kawasaki}}, \bibinfo {author} {\bibfnamefont {Yutaka}\ \bibnamefont
  {Kishimoto}}, \bibinfo {author} {\bibfnamefont {Takashi}\ \bibnamefont
  {Ohno}}, \bibinfo {author} {\bibfnamefont {Yoshitaka}\ \bibnamefont
  {Michihiro}}, \bibinfo {author} {\bibfnamefont {Zhangzhen}\ \bibnamefont
  {He}}, \bibinfo {author} {\bibfnamefont {Yutaka}\ \bibnamefont {Ueda}}, \
  and\ \bibinfo {author} {\bibfnamefont {Mitsuru}\ \bibnamefont {Itoh}},\
  }\bibfield  {title} {\enquote {\bibinfo {title} {{${}^{51}$V NMR study of
  antiferromagnetic state and spin dynamics in quasi-one-dimensional
  BaCo${}_{2}$V${}_{2}$O${}_{8}$}},}\ }\href {\doibase
  10.1103/PhysRevB.86.094433} {\bibfield  {journal} {\bibinfo  {journal} {Phys.
  Rev. B}\ }\textbf {\bibinfo {volume} {86}},\ \bibinfo {pages} {094433}
  (\bibinfo {year} {2012})}\BibitemShut {NoStop}%
\end{thebibliography}

\end{document}